\documentclass[
prb,twocolumn,
superscriptaddress,
amsmath,amssymb,
aps
]{revtex4-2}

\usepackage{graphicx}
\usepackage{dcolumn}
\usepackage{bm}
\usepackage[dvipsnames]{xcolor}
\usepackage{hyperref}
\usepackage{mathtools}
\usepackage{tabularx}

\DeclarePairedDelimiter{\ceil}{\lceil}{\rceil}
\hypersetup{
    colorlinks=true,
    linkcolor=MidnightBlue,
    citecolor=RedViolet,
    filecolor=RedViolet,      
    urlcolor=RedViolet,
    }

\newcommand{\veryshortarrow}[1][3pt]{\mathrel{%
   \hbox{\rule[\dimexpr\fontdimen22\textfont2-.2pt\relax]{#1}{.4pt}}%
   \mkern-7mu\hbox{\usefont{U}{lasy}{m}{n}\symbol{40}}}}
   
\begin{document}

\title{Competing instabilities of the extended Hubbard model on the triangular lattice:\\ 
Truncated-unity functional renormalization group and application to moir\'e materials}

\author{Nico Gneist}
\affiliation{Institut f\"ur Theoretische Physik, Universit\"at zu K\"oln, D-50937 Cologne, Germany}

\author{Laura Classen}
\affiliation{Max Planck Institute for Solid State Research, D-70569 Stuttgart, Germany}%

\author{Michael M. Scherer}
\affiliation{Institut f\"ur Theoretische Physik III, Ruhr-Universit\"at Bochum, D-44801 Bochum, Germany}

\date{\today}

\begin{abstract}
A simple yet paradigmatic model for the interplay of strong electronic correlations and geometric frustration is the triangular lattice Hubbard model.
Recently it was proposed that moir\'e structures of transition metal dichalcogenides can be used to simulate extended versions that include non-local density-density interactions.
We study competing instabilities of interacting electrons in such an extended Hubbard model
on the triangular lattice near a filling where the density of states has a Van Hove singularity.
We employ a truncated-unity functional renormalization group approach to investigate two cases: a paradigmatic minimally extended Hubbard model and a specific model with parameters that are applicable to hetero-bilayers of transition metal dichalcogenides.
We unravel rich phase diagrams, including tendencies to spin-density-wave order and unconventional pairing, which can give rise to  topological superconductivity.
We classify the symmetry of the superconducting instabilities according to their irreducible representations and show that higher lattice harmonics are dominant when the nearest-neighbor interaction is sizable indicating pair formation between second-nearest neighbors.
The phenomenological consequences can be enhanced spin and thermal quantum Hall responses in a topological superconductor.
\end{abstract}

\maketitle

\section{ Introduction}\label{sec:intro}

The triangular lattice Hubbard model is a  paradigmatic model for the study of the complex interplay between strong electronic correlations and geometric frustration.
A very diverse set of phenomena has been associated with its phase diagram including metal-insulator transitions, various magnetic states, chiral superconductivity, and quantum spin liquid phases~\cite{arovas2021hubbard,qin2021hubbard}.
The advent of moir\'e materials~\cite{cao2018magic,cao2018correlated,yankowitz2019tuning,kerelsky2019maximized,sharpe2019emergent,lu2019superconductors,serlin2020intrinsic} opened new opportunities for the study of such correlation phenomena because they can be engineered to be quantum simulators for many models of strongly correlated electron systems~\cite{kennes2021moire}. 
In particular, it has been argued that moir\'e hetero-bilayers of transition metal dichalcogenides (TMD) can be used to simulate the single-band triangular-lattice Hubbard model~\cite{PhysRevLett.121.026402}.  
For example, in the hetero-bilayer system WSe${}_2$/MoS${}_2$, a moir\'e potential is induced by a mismatch in orientation and a difference in lattice constants. 
This moir\'e potential was predicted to modify the top spin-split valence band of WSe${}_2$ in such a way that it gives rise to an isolated, narrow band close to the Fermi level, whose filling and interactions can be controlled by gating and screening layers~\cite{PhysRevLett.121.026402}.

\begin{figure}[t!]
\includegraphics[width=\columnwidth]{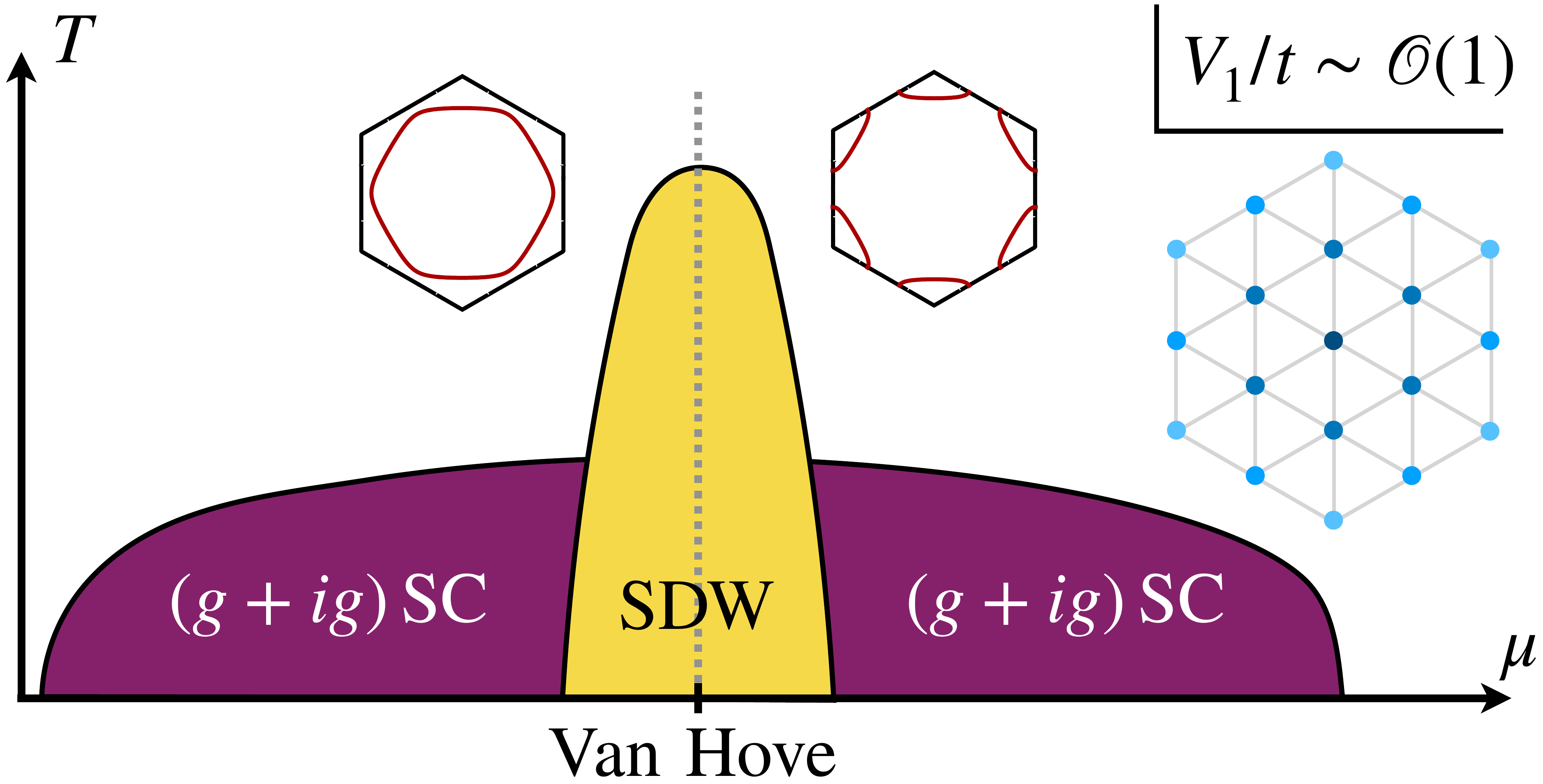}
\caption{\textbf{Schematic phase diagram} of the extended Hubbard model on the triangular lattice.
The nearest-neighbor repulsion $V_1$, which is present in addition to a moderate local Hubbard term $U/V_1\sim 4$, supports the formation of a $(g+ig)$-wave superconducting state.
This state is topologically non-trivial and features enhanced quantum Hall responses as compared to $(d+id)$ superconductivity.
Quantitative, extended versions of this phase diagram are shown in Figs.~\ref{fig:PhaseDiagram1}, \ref{fig::convergence1092} and \ref{fig:PhaseDiagram2} for a minimal model and for an accurate fit of the AA stacked moir\'e hetero-bilayer TMD WSe${}_2$/MoS${}_2$.}
\label{fig:schematic}
\end{figure}

A first generation of experiments on 
moir\'e hetero-bilayer TMDs~\cite{regan2020mott,tang2020simulation,jin2021stripe,li2021continuous,huang2021correlated} revealed a rich phenomenology associated with strong correlation effects: 
the occurrence of Mott insulating behavior, a continuous metal-insulator transition, antiferromagnetic and ferromagnetic responses at commensurate fillings, and generalized Wigner crystallization.
Furthermore, a relatively sharp drop of the resistivity has been observed in a closely related material system, the moir\'e homo-bilayer WSe${}_2$/WSe${}_2$, which has been interpreted as a sign for a possible superconducting state~\cite{wang2020correlated}.

On the theory side, a series of sophisticated numerical 
studies of the triangular-lattice Hubbard model has recently been put forward focusing on the half-filled case~\cite{PhysRevB.94.245145,PhysRevX.10.021042,PhysRevB.102.115136,PhysRevB.103.235132,chen2021quantum,PhysRevX.11.041013} and slightly doping away from it~\cite{zhu2020doped,PhysRevB.103.165138}. 
In addition, the important effect of extended Hubbard interactions, which are sizable in moir\'e TMDs, was investigated~\cite{PhysRevB.102.201104,zhou2021quantum,morales2021non}.  
Away from half filling, several studies have discussed indications for topological superconducting phases, which are facilitated in this system due to the hexagonal lattice structure~\cite{PhysRevB.68.104510,nandkishore2012chiral,PhysRevB.88.041103,PhysRevB.91.245125,PhysRevB.89.144501,PhysRevB.100.060506,huang2021topological,PhysRevB.81.224505,PhysRevLett.111.097001,PhysRevB.98.174515,wolf2021triplet,scherer2021mathcal}, and are also strongly affected by the presence of extended Hubbard interactions~\cite{PhysRevB.97.155145,PhysRevLett.110.166401,scherer2021mathcal,PhysRevB.98.174515,wolf2021triplet,PhysRevLett.111.097001}.

Here, we explore the triangular-lattice Hubbard model away from half filling including extended Hubbard interactions and expand upon previous studies utilizing the truncated unity functional renormalization group (TUFRG)~\cite{lichtenstein2017high}, see also Refs.~\cite{PhysRevB.96.205155,PhysRevB.95.085143,PhysRevB.96.205155,PhysRevB.103.235150,PhysRevResearch.3.023180}.
We specifically focus on the case near Van-Hove filling where the model exhibits a singularity in the density of states and investigate two model cases: 
the minimal, paradigmatic $t-U-V$ case with only nearest-neighbor hopping, on-site and nearest-neighbor interaction, and the WSe$_2$/MoS$_2$ case with parameter values from \textit{ab initio} modelling~\cite{PhysRevLett.121.026402} for up to the third-neighbor hopping and interaction. 
In both cases we provide comprehensive phase diagrams of Fermi-surface instabilities for varying filling and strength of extended interactions. 
We obtain an instability towards a spin density wave at Van Hove filling due to perfect (approximate) nesting in the $t-U-V$ (WSe$_2$/MoS$_2$) case, and a rich structure of pairing instabilities at its low and high doping side. 
In particular, we find that the inclusion of extended, repulsive interactions yields pairing states with dominant second-nearest-neighbor contributions that belong to the 2D irreducible representation $E_2$ of the lattice symmetry group $D_6$. 
This indicates the possible formation of a topological $g+ig$ superconducting state.
We argue that it is crucial to account for the coupling between ordering tendencies for an unbiased analysis and to accurately resolve the momentum-dependence for the determination of pairing symmetries, both of which are strong suits of the TUFRG.

We introduce the models in Sec.~\ref{sec:model} and our method in Sec.~\ref{sec:method}.
Results for the correlated states of the minimal model are presented in Sec.~\ref{sec:results} and for the moir\'e TMD model in Sec.~\ref{sec:results2}.
We discuss the possible topological superconductivity in Sec.~\ref{sec:g+ig} and conclude in Sec.~\ref{sec:conclusion}.

\section{Model}\label{sec:model}

We consider 
an extended Hubbard model on the triangular-lattice with SU(2)-spin symmetry 
of the form 
\begin{align}
    H =& -\sum_{n}\sum_{\langle ij\rangle_n}\sum_\sigma t_n\left(c^\dagger_{i\sigma}c_{j\sigma}+ \mathrm{h.c.}\right)-\mu \sum_{i\sigma}n_{i\sigma}\nonumber\\
    &+U\sum_{i}n_{i\uparrow}n_{i\downarrow}+\sum_{n}\sum_{\langle ij\rangle_n }\sum_{\sigma\sigma'} V_n n_{i\sigma}n_{j\sigma'}\,.
    \label{eq:model}
\end{align}
In the first line, we have introduced the electron annihilation (creation) operators $c^{(\dagger)}_{i\sigma}$ for lattice site $i$, spin projection $\sigma \in \{\uparrow,\downarrow\}$, and the
$n$-th neighbor hopping $t_n$ denoted by $\langle i,j\rangle_n$ in the sum.
The electron density operator is $n_{i\sigma} = c^\dagger_{i\sigma}c_{i\sigma}$ and couples to the chemical potential $\mu$ which allows us to adjust the electron filling of the system. 
In the second line of Eq.~\eqref{eq:model}, we collect interaction terms, i.e. the onsite Hubbard interaction with interaction parameter $U$ and the 
n-th neighbor Coulomb interaction $V_n$.
We study two specifications of the model in Eq.~\eqref{eq:model}. 
To investigate the effect of extended interactions in a clear setup, we consider the minimal model with $t_1=t>0$, $U>V_1>0$ and all other $t_n=0=V_n$ ($n\geq2$). 
For a realistic example, we consider parameters for the highest moir\'e valence band of the heterobilayer WSe$_2$/MoS$_2$ with $t_1\approx -2.5\,\mathrm{meV}, t_2\approx 0.5\,\mathrm{meV}, t_3\approx 0.25\,\mathrm{meV}$ and up to the third-nearest-neighbor interaction $U>V_1>V_2>V_3>0$.
Farther-ranged hoppings and interactions $n\geq4$ are again set to zero. 
Note that $\sigma$ in the effective triangular-lattice Hubbard model for hetero-bilayer TMD describes a pseudo-spin corresponding to a spin-valley locked degree of freedom.

\begin{figure}[t!]
\includegraphics[width=0.95\columnwidth]{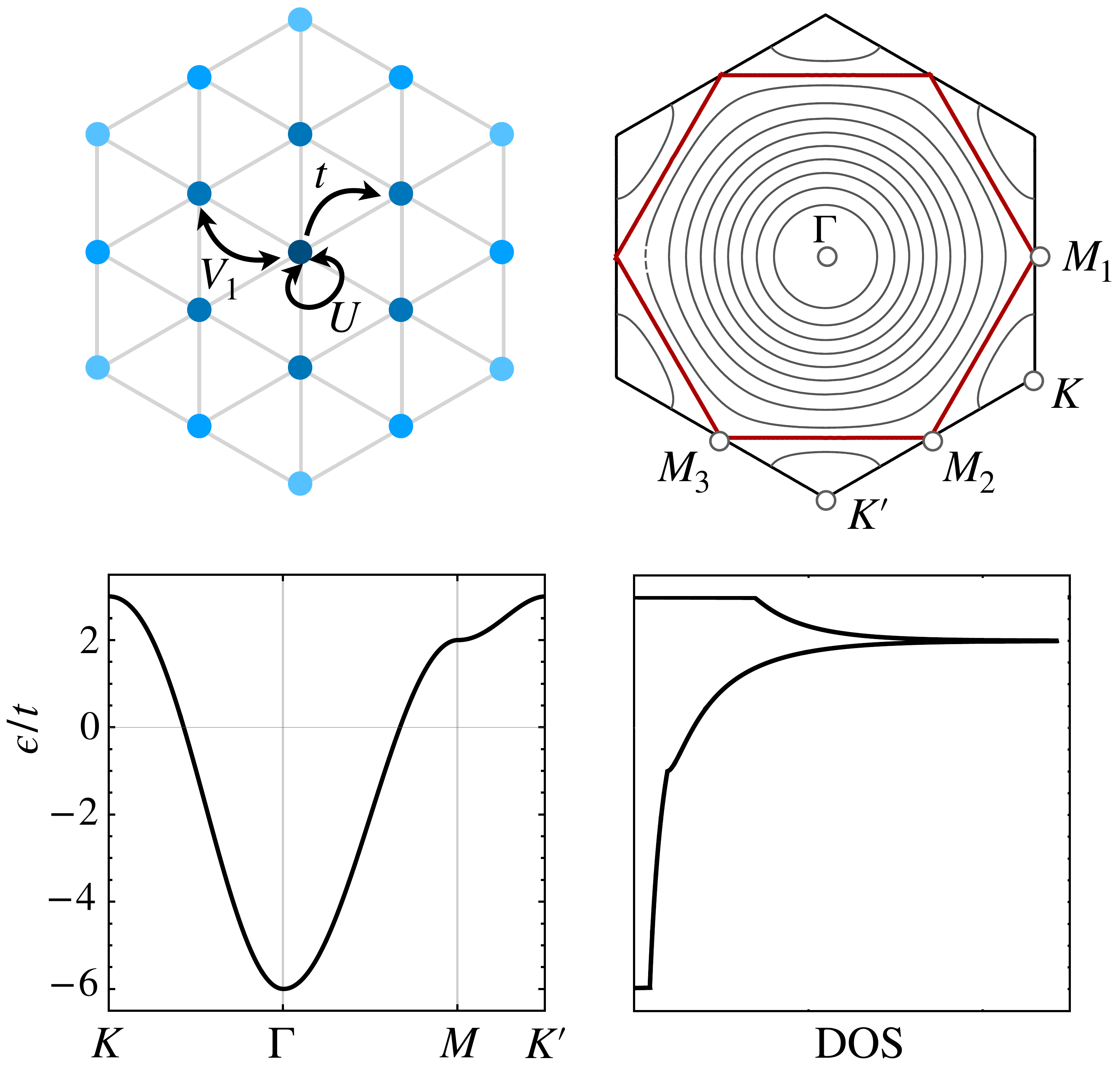}
\caption{\textbf{Paradigmatic model and energy dispersion.} Top left: Triangular lattice and model parameters. Top right: BZ and high-symmetry points with energy contours. The Fermi line for the case of Van-Hove filling exhibits perfect nesting and is shown as the thick red line. Bottom left: energy dispersion along a path in the BZ. Bottom right: density of states.}
\label{fig:model}
\end{figure}

In the $t-U-V$ case, we obtain the energy dispersion
\begin{align}\label{eq:dispersion}
   \xi(\bm{k})\!=\!-2t[\cos(k_x)\!+\!2\cos(k_x/2)\cos(\sqrt{3}k_y/2)]-\mu\,,
\end{align}
with wavevector $\bm{k}=(k_x,k_y)$ from the first line of Eq.~\eqref{eq:model}, i.e. the tight-binding part, cf. Fig.~\ref{fig:model}.
There is a Van-Hove singularity in the density of states (DOS) at $\mu=2t$, which is caused by saddle points in the energy dispersion that occur at the three inequivalent $M_i$ points. 
At Van Hove filling, the Fermi surface is also perfectly nested.
In the WSe$_2$/MoS$_2$ case, the dispersion is very similar
\begin{align}\label{eq:dispersion2}
   \xi(\bm{k})=&-2t_1[\cos(k_x)\!+\!2\cos(k_x/2)\cos(\sqrt{3}k_y/2)] \notag \\
   &-2t_2[2\cos(3k_x/2)\cos(\sqrt{3}k_y/2)+\cos(\sqrt{3}k_y)] \notag \\
   &+2t_3[\cos(2k_x)-2\cos(k_x)\cos(\sqrt{3}k_y)] -\mu\,.
\end{align}
Note, however, that it is inverted in comparison to Fig.~\ref{fig:model}, i.e. hole-like, since $t_1<0$. The Van Hove singularity appears at $\mu=2t_1+2t_2-6t_3$, and nesting is still a very good approximation. 

\begin{figure}[t!]
\includegraphics[width=0.82\columnwidth]{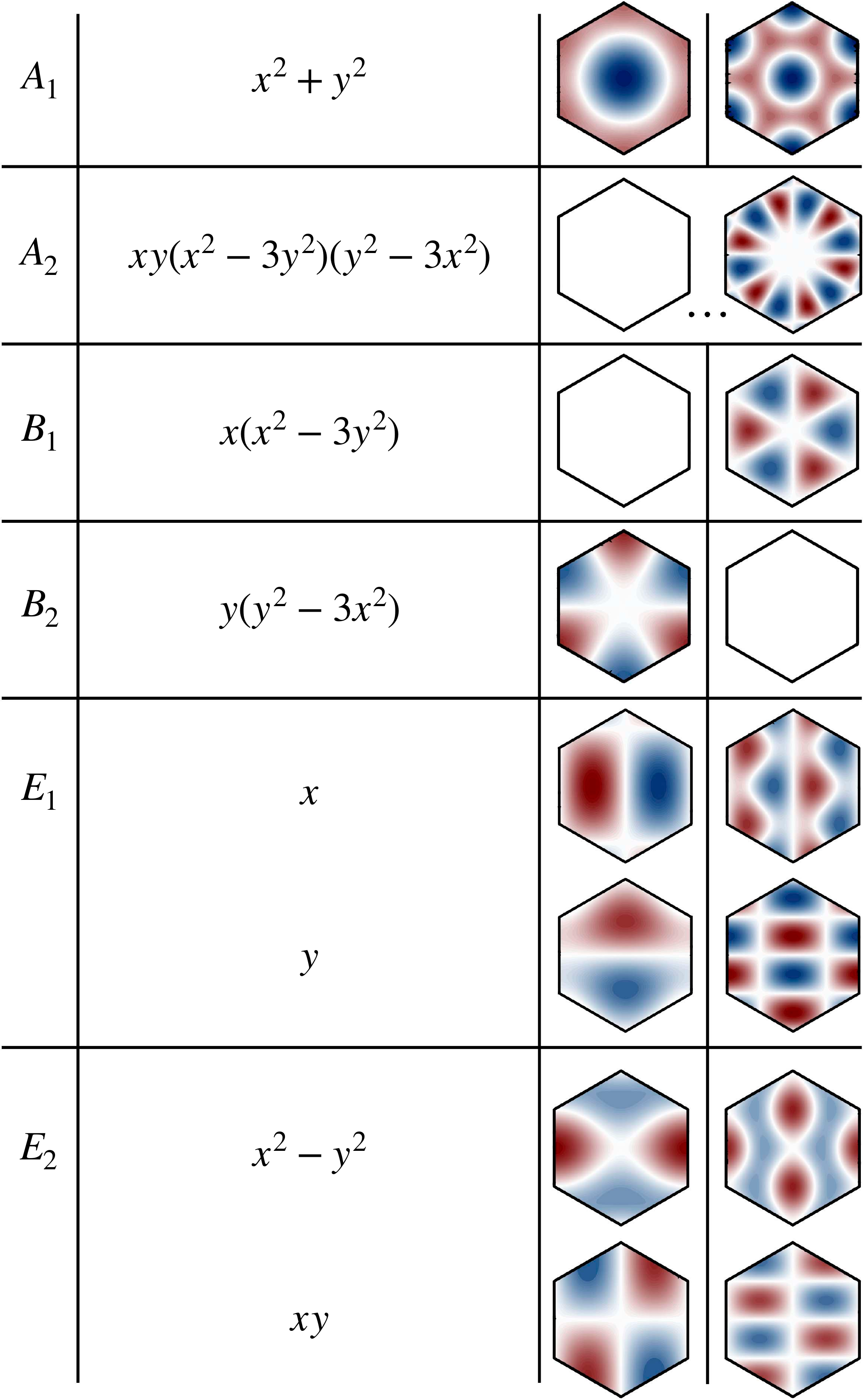}
\caption{\textbf{Lattice symmetry group $D_6$.} 
First column: irreducible representations
Second column: basis function(s).
Third column: nearest-neighbor lattice harmonic(s).
Fourth column: second-nearest-neighbor lattice harmonic(s). 
Exception in the second row ($A_2$): 
a non-zero lattice harmonic only occurs at higher order as indicated by the ellipsis.}
\label{fig:harmony}
\end{figure}

For later reference, we also recall some properties of the point group of the triangular lattice, i.e. the dihedral group $D_6$. 
This symmetry group has four one-dimensional irreducible representations (irreps), $A_1$, $A_2$, $B_1$, and $B_2$, and two two-dimensional irreps, $E_1$ and $E_2$. 
We depict the corresponding lowest-order non-vanishing basis functions with their nearest- and second-nearest-neighbor lattice harmonics in Fig.~\ref{fig:harmony}.
Below, we will use these irreps and lattice harmonics to classify the superconducting pairing vertex in detail, which allows us to extract the symmetry properties of the superconducting gap function that would develop out of an instability of the pairing vertex. 
The gap function must be totally antisymmetric under electron exchange and it involves a wave-vector-dependent part and a spin-dependent part, i.e. it comes as spin-singlet or spin-triplet pairing depending on the transformation of the corresponding irrep under parity.
The irreps $A_1, A_2, E_2$ are even under parity, i.e. they can describe spin-singlet superconducting states, and the irreps $B_1, B_2, E_1$ are odd under parity, i.e. they can describe spin-triplets.

\section{Method}\label{sec:method}

The understanding of strongly-correlated electron systems relies on the development of appropriate quantum many-body methods.
A versatile and powerful approach to strongly-correlated electrons, which can generally cover  a broad parameter range (e.g. for band structure, filling, interaction types) is the functional renormalization group (FRG)~\cite{WETTERICH199390,Dupuis:2020fhh,RevModPhys.84.299,platt2013functional}.
The FRG uses functional methods within the framework of quantum field theory and its broad applicability comes at the cost of introducing truncation schemes.
In the context of strongly-correlated electrons, one profitable truncation focuses on the evolution of the two-particle vertex and provides a tool to identify the leading Fermi-surface instabilities in the presence of competing interactions~\cite{RevModPhys.84.299,platt2013functional}.
More sophisticated truncation schemes or a combination with other powerful many-body methods also provide means to go beyond such instability analyses~\cite{PhysRevB.70.115109,PhysRevB.70.125111,PhysRevB.79.094526,PhysRevB.81.235108,PhysRevB.83.155125,PhysRevB.87.174523,PhysRevB.89.035126,PhysRevB.89.121116,PhysRevLett.112.196402,PhysRevLett.116.096402,PhysRevLett.120.057403,PhysRevB.97.035162,10.21468/SciPostPhys.6.1.009,PhysRevResearch.2.033372,PhysRevResearch.4.013034}.

In practice, one starts the RG procedure from an initial ultraviolet renormalization group scale $\Lambda_0$, which corresponds to the bandwidth of the system and then successively integrates out electron fluctuations in a Wilsonian-like RG scheme to the infrared scale $\Lambda \to 0$, where all fluctuations are included.
The method has been successfully applied to correlated electrons on hexagonal lattices revealing complex phase diagrams in, e.g., paradigmatic models~\cite{PhysRevB.68.104510,honerkamp2008density}, single- and multi-layer graphene~\cite{kiesel2012competing,PhysRevB.85.035414,PhysRevB.87.094521,scherer2012instabilities,scherer2012interacting,PhysRevB.90.035122}, various unconventional superconductor 
candidates~\cite{PhysRevLett.111.097001,PhysRevB.89.020509,2015NatCo.6.8232E,wolf2021triplet}, and, more recently, correlated moir\'e materials~\cite{PhysRevB.98.241407,PhysRevB.99.094521,PhysRevB.99.195120,scherer2021mathcal}.

\subsection{FRG background}\label{sec:basicsFRG}

We start with the action for a many-electron system as described by a corresponding Hamiltonian, cf. Eq.~\eqref{eq:model}, 
%
\begin{align}\label{eq:action}
	S[\bar\psi,\psi]=-(\bar\psi,G_0^{-1}\psi)+V[\bar\psi,\psi]\,,
\end{align}
where $\bar\psi,\psi$ are Grassmann-valued electron fields. 
The first term is quadratic in the electron fields and includes the free  propagator $G_0(\omega,\bm{k})=1/(i\omega-\xi({\bm{k}}))$ with Matsubara frequency $\omega$
and the bracket~$(.,.)$ denotes integration over frequencies, wave-vectors, and further quantum numbers.
The second term $V[\bar\psi,\psi]$ in Eq.~\eqref{eq:action} is an interaction term which is quartic in the electron fields. 
It can be read off from the interaction part of Eq.~\eqref{eq:model}.

To set up the renormalization group scheme, the free propagator is regularized by an infrared cutoff $\Lambda$, i.e.
\begin{align}
	G_0(\omega,\bm{k})\to G_0^\Lambda(\omega,\bm{k})\,.
\end{align}
This regularization procedure cuts off the infrared modes below the scale $\Lambda$ and its concrete implementation can be done in different ways. 
We choose the temperature as the scale parameter in our calculations below, see App.~\ref{sec:regulator}, but in this section we leave the choice open as it does not affect the general structure of the FRG equations.
The effective action $\Gamma[\bar\psi,\psi]$, generating the one-particle irreducible correlation functions~\cite{negele2018quantum}, is then defined with $G_0^\Lambda$ and hence becomes scale dependent, i.e. $\Gamma \to \Gamma^\Lambda$.

Taking the derivative of $\Gamma^\Lambda$ with respect to $\Lambda$ yields the exact RG flow equation~\cite{RevModPhys.84.299}
\begin{align}
	\frac{\partial}{\partial \Lambda}\Gamma^\Lambda\!=\!-(\bar\psi,(\dot{G}^\Lambda_0)^{-1}\psi)\!-\!\frac{1}{2}\mathrm{Tr}\left(\!(\dot{\mathbf{G}}^\Lambda_0)^{-1}
	(\mathbf{\Gamma}^{(2)\Lambda})^{-1}
	\!\right),\label{eq:exRG}
\end{align}
where $(\mathbf{G}^\Lambda_0)^{-1}=\mathrm{diag}((G^\Lambda_0)^{-1},(G^{\Lambda t}_0)^{-1})$ and the trace includes the matrix of second functional derivatives of $\Gamma^\Lambda$ with respect to $\psi$ and $\bar\psi$, i.e. $\mathbf{\Gamma}^{(2)\Lambda}=\mathbf{\Gamma}^{(2)\Lambda}[\bar\psi,\psi]$. 
The inital condition of Eq.~\eqref{eq:exRG} is defined at ultraviolet scale $\Lambda_0$ and corresponds to the microscopic action $\Gamma^{\Lambda_0}\!=\!S$. 
For $\Lambda\!\to\! 0$ the full effective action 
is restored $\Gamma^{\Lambda\rightarrow 0}=\Gamma$.

For practical calculations on the basis of Eq.~\eqref{eq:exRG}, we need to truncate $\Gamma^\Lambda$.
The truncation employed here is based on a vertex expansion, reading
\begin{align}
	\Gamma^\Lambda[\psi,\bar\psi]=&\sum_{i=0}^\infty \frac{(-1)^i}{(i!)^2}\!\!\sum_{\substack{K_1,...,K_i\\ K_1^\prime,..., K_i^\prime}}\!\!\!\Gamma^{(2i)\Lambda}(K_1^\prime,..., K_i^\prime,K_1,...,K_i)\nonumber\\[3pt]
	&\times \bar\psi(K_1^\prime)...\bar\psi(K_i^\prime)\psi(K_i)...\psi(K_1)\,,
\end{align}
where $K=(\sigma,k)$ carries spin indices $\sigma$ and multi-indices $k =(\omega,\vec k)$ collecting Matsubara frequencies and wavevectors. 
This ansatz is inserted into Eq.~\eqref{eq:exRG}, generating a hierarchy of RG flow equations for the one-particle irreducible vertex functions $\Gamma^{(2i)\Lambda}$.
Following earlier work, we truncate the tower of flow equations at the second level $i=2$ and neglect self-energy feedback~\cite{salmhofer2001fermionic,PhysRevB.67.174504} so that we exclusively consider the RG evolution of the two-particle vertex $\Gamma^{(4)\Lambda}$, which determines the effective interaction dressed by multiple scattering events.
This truncation scheme accurately resolves the wave-vector dependence of the two-particle vertex, which allows us to determine Fermi liquid instabilities in an unbiased way.

\subsection{Spin-invariant FRG flow equations}\label{sec:SU2}
 
We can utilize the spin invariance of our model Eq.~\eqref{eq:model} in the FRG equation for the two-particle vertex.
For SU(2) symmetric systems it can be written as
\begin{align}\label{eq::Gamma4_0}
	\Gamma^{(4)\Lambda}_{\sigma_1\sigma_2\sigma_3\sigma_4}=V^\Lambda\delta_{\sigma_1\sigma_3}\delta_{\sigma_2\sigma_4}-\tilde V^\Lambda\delta_{\sigma_1\sigma_4}\delta_{\sigma_2\sigma_3}\,,
\end{align}
with effective interaction 
$V^\Lambda=V^\Lambda(k_1,k_2,k_3,k_4)$ and $\tilde V^\Lambda=V^\Lambda(k_1,k_2,k_4,k_3)$.
For later convenience, we also explicitly introduce the effective  interaction as it appears in the action, i.e.
\begin{align}
\Gamma^{\Lambda}_V=&\frac{1}{2}\int_{k_1,k_2,k_3,k_4}\!\! V^{\Lambda}(k_1,k_2,k_3,k_4) \delta(k_1\!+\!k_2\!-\!k_3\!-\!k_4)\nonumber\\[5pt]
&\times \sum_{s,s'} \bar{\psi}_s(k_1) \bar{\psi}_{s'}(k_2) \psi_{s'}(k_4) \psi_{s}(k_3)\,,
\label{eq::Gamma4}
\end{align}
where $\int_k= A_{\mathrm{BZ}}^{-1}T\int_{\mathrm{BZ}} d\bm{k} \sum_{i\omega}$ and $k=(\bm{k},\omega)$. $A_{\mathrm{BZ}}$ is the area of the BZ. 
In the following, we omit the fourth momentum argument which is fixed by momentum conservation unless it is important for the discussion.

For the investigation of Fermi-surface instabilities, we are interested in analyzing the most singular part of $V^\Lambda$, which comes from the smallest Matsubara frequency and we therefore exclusively consider this one.
With these preliminaries, the RG evolution of $V^\Lambda$ can be derived from the exact flow equation, Eq.~\eqref{eq:exRG}, and it reads
\begin{align}\label{eq:flowequation}
	\frac{d}{d\Lambda}V^\Lambda=\tau_\mathrm{pp}+\tau_\mathrm{ph,c}+\tau_\mathrm{ph,d}\,.
\end{align}
This RG flow equation is composed of three contributions, i.e. the particle-particle (pp) contribution
\begin{align}
    \tau_{\mathrm{pp}}=&-\!\int_q  \frac{d}{d\Lambda}[G_0^\Lambda(i\omega, \bm{q}+\bm{k}_1+\bm{k}_2)G_0^\Lambda(-i\omega,-\bm{q})] \nonumber\\[3pt]
    & \quad\times V^{\Lambda}(\bm{k_1},\bm{k_2},\bm{q}+\bm{k_1}+\bm{k_2})\nonumber\\[3pt]
    &\quad\times  V^{\Lambda}(\bm{q}+\bm{k}_1+\bm{k}_2,-\bm{q},\bm{k}_3)\,,\label{eq::contribution1}
\end{align}
the crossed particle-hole (ph,c) contribution
\begin{align}
    \tau_{\mathrm{ph,c}}=&-\!\int_q \frac{d}{d\Lambda}[G_0^\Lambda(i\omega, \bm{q}+\bm{k}_1-\bm{k}_4)G_0^\Lambda(i\omega,\bm{q})] \nonumber\\[3pt]
    & \quad\times V^{\Lambda}(\bm{k}_1,\bm{q},\bm{q}\!+\!\bm{k}_1\!-\bm{k}_4) \nonumber\\[3pt]
    &\quad\times V^{\Lambda}(\bm{q}+\bm{k}_1-\bm{k}_4,\bm{k}_2,\bm{k}_3)\,,\label{eq::contribution2}
\end{align}
and the direct particle-hole (ph,d) contribution
\begin{align}
    \tau_{\mathrm{ph,d}} =&-\!\int_q \frac{d}{d\Lambda}[G_0^\Lambda(i\omega, \bm{q}+\bm{k}_1-\bm{k}_3)G_0^\Lambda(i\omega,\bm{q})] \nonumber\\[3pt]
     &\ \bigl[\ V^{\Lambda}(\bm{k}_1,\bm{q},\bm{q}\!+\!\bm{k}_1\!-\!\bm{k}_3) V^{\Lambda}(\bm{q}\!+\!\bm{k}_1\!-\!\bm{k}_3,\bm{k}_2,\bm{q}) \nonumber\\[3pt]
&+ V^{\Lambda}(\bm{k}_1,\bm{q},\bm{k}_3) V^{\Lambda}(\bm{q}\!+\!\bm{k}_1\!-\!\bm{k}_3,\bm{k}_2,\bm{k}_4)\nonumber\\[3pt]
&-2V^{\Lambda}(\bm{k}_1,\bm{q},\bm{k}_3)
 V^{\Lambda}(\bm{q}\!+\!\bm{k}_1\!-\!\bm{k}_3,\bm{k}_2,\bm{q})\ \bigr]\,.\label{eq::contribution3}
\end{align}
By solving the flow equation~\eqref{eq:flowequation} using  Eqs.~\eqref{eq::contribution1}--\eqref{eq::contribution3} we can identify Fermi-surface instabilities in terms of singular contributions to $V^{\Lambda}(\bm{k}_1,\bm{k}_2,\bm{k}_3)$.
We use this set of equations as the basis for an implementation of the computationally efficient TUFRG scheme.

\subsection{Truncated-Unity FRG}\label{sec:tufrg}

\begin{figure}[t!]
\includegraphics[width=\columnwidth]{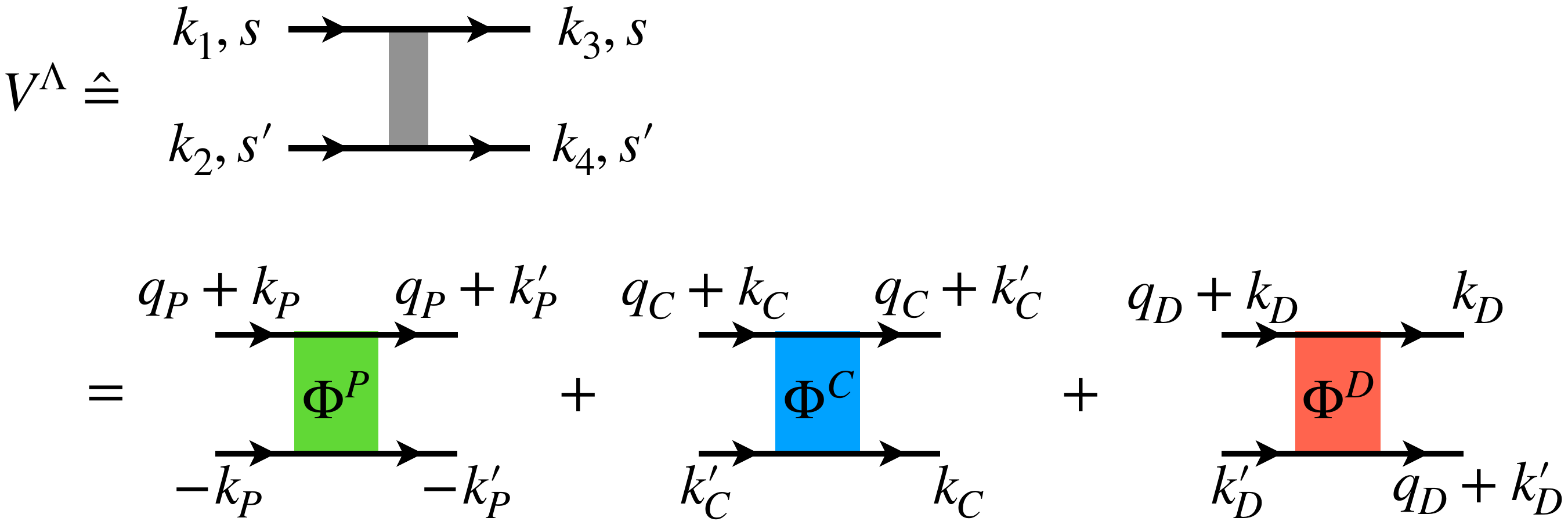}\\[3pt]
\begin{tabular}{p{105pt}|p{40pt}|p{40pt}|p{40pt}}
\hline\hline
 Channel $X$ &  $P$& $C$& $D$\\
 \hline
 Interaction type & Pairing & Magnetic & Density\\[3pt]
 Transfer momentum $q_X$ &$\bm{k}_1+\bm{k}_2$ &$\bm{k}_1-\bm{k}_4$ &$\bm{k}_1-\bm{k}_3$  \\[3pt]
 Momentum $k_X$ & $-\bm{k}_2$ & $\bm{k}_4$ &  $\bm{k}_3$ \\[3pt]
 Momentum $k'_X$ & $-\bm{k}_4$ & $\bm{k}_2$ &  $\bm{k}_2$ \\[3pt]
 Flow contribution &$\tau_{\mathrm{pp}}$ & $\tau_{\mathrm{ph,c}}$&$\tau_{\mathrm{ph,d}}$\\
 \hline\hline
\end{tabular}
\caption{\textbf{Channel decomposition of the vertex} $V^{\Lambda}$. 
By explicitly inserting the decomposition Eq.~\eqref{eq::decomposition} into the vertex Eq.~\eqref{eq::Gamma4} and relabeling the important momentum as $q_X$ and the two remaining momenta as $k_X, k'_X$, $X \in \{P,C,D\}$, the channels can be associated with  superconducting, magnetic and density fluctuations. Spin indices are as in the first line.}
\label{fig:decompo}
\end{figure}

The singular behavior of Fermi-surface instabilities in correlated-electron systems typically occurs in the transfer momenta in the three loop contributions in Eq.~\eqref{eq:flowequation}~\cite{PhysRevB.79.195125}. 
To facilitate a high resolution of the transfer momenta, we introduce a  singular-momentum description of the RG evolution equations reparameterizing the vertices in Eqs.~\eqref{eq::contribution1}--\eqref{eq::contribution3} by introducing different interaction channels.
To that end, $V^{\Lambda}$ is decomposed as
\begin{align}
    V^{\Lambda}(\bm{k}_1,\bm{k}_2,\bm{k}_3,\bm{k}_4)=& \ V^{\Lambda,0}(\bm{k}_1,\bm{k}_2,\bm{k}_3,\bm{k}_4)  \nonumber\\[3pt] 
    &+\Phi^{\Lambda,P}(\bm{k}_1+\bm{k}_2;-\bm{k}_2,-\bm{k}_4)   \nonumber\\[3pt] &+\Phi^{\Lambda,C}(\bm{k}_1-\bm{k}_4;\bm{k}_4,\bm{k}_2)  \nonumber\\[3pt]
    &+\Phi^{\Lambda,D}(\bm{k}_1-\bm{k}_3;\bm{k}_3,\bm{k}_2)\,,
    \label{eq::decomposition}
\end{align}
where $V^{\Lambda,0}(\bm{k}_1,\bm{k}_2,\bm{k}_3,\bm{k}_4)$ takes care of the initial condition. 
The other three contributions $\Phi^{\Lambda,X}$ with $X \in \{P,C,D\}$ are the actual channels and in each case the transfer momentum is the first argument, see Fig.~\ref{fig:decompo}. 
Inserting Eq.~\eqref{eq::decomposition} into Eq.~\eqref{eq::Gamma4} and rearranging the terms, we find that each of these three channels describes a specific physical interaction, see also Fig.~\ref{fig:decompo}.

These channels can be defined via their respective RG contributions in Eqs.~\eqref{eq::contribution1}--\eqref{eq::contribution3}, i.e., via the three flow equations
\begin{align}
    \frac{d}{d\Lambda} \Phi^{P}(\bm{k}_1+\bm{k}_2;-\bm{k}_2,-\bm{k}_4) &= \tau_{\mathrm{pp}}(\bm{k}_1,\bm{k}_2,\bm{k}_3,\bm{k}_4)\label{eq::diff3a}\,,  \\[3pt]
    \frac{d}{d\Lambda} \Phi^{C}(\bm{k}_1+\bm{k}_4;\bm{k}_4,\bm{k}_2) &= \tau_{\mathrm{ph,c}}(\bm{k}_1,\bm{k}_2,\bm{k}_3,\bm{k}_4)\,,  \label{eq::diff3b} \\[3pt]
    \frac{d}{d\Lambda} \Phi^{D}(\bm{k}_1-\bm{k}_3;\bm{k}_3,\bm{k}_2) &= \tau_{\mathrm{ph,d}}(\bm{k}_1,\bm{k}_2,\bm{k}_3,\bm{k}_4)\label{eq::diff3c}\,,
\end{align}
where we have dropped the index~$\Lambda$ for convenience.
The first wave-vector argument in Eqs.~\eqref{eq::diff3a}--\eqref{eq::diff3c} labels the transfer momentum.
The dependence on the other two momenta can be expanded in a form-factor basis 
\begin{equation}
    \Phi^{X}(\bm{q},\bm{k},\bm{k}') = \sum_{l,l'} X^{l,l'}(\bm{q}) f_l(\bm{k}) f_{l'}^{*}(\bm{k}')\,.\label{eq::formfactorexpansion}
\end{equation}
The above expansion holds for form factors forming a unity with respect to $l$ and $\bm{k}$
\begin{align}
    A_{\mathrm{BZ}}^{-1}\sum_{l} f_{l}^{*}(\bm{p})f_{l}(\bm{k})&= \delta(\bm{p}-\bm{k})\,, \label{eq::unity1}\\[3pt]
    A_{\mathrm{BZ}}^{-1}\int d\bm{k} f_{l}^{*}(\bm{k})f_{l'}(\bm{k})&= \delta_{l,l'}\,.
    \label{eq::unity}
\end{align}
In the numerical implementation this expansion is truncated, i.e. the $l,l'$ sum is restricted to a finite number $N_l$ of form factors to resolve the weaker momentum dependence in $\bm{k}$ and $\bm{k}'$.
In contrast, the transfer momentum $\bm{q}$ carrying a strong momentum dependence is discretized in a momentum mesh in the BZ with resolution $N_{\bm{q}}$.

The above decomposition of the vertex can be turned into a computational advantage.
This is because the first description of the vertex $V^{\Lambda}(\bm{k}_1,\bm{k}_2,\bm{k}_3)$ using a wavevector resolution $N_{\bm{k}}$ of the BZ, leads to a set of $N_{\bm{k}}^3$ coupled differential equations. 
In contrast, the channel-decomposed vertices scale with $\propto N_{\bm{q}}\times N_l^2$. 
Therefore, by truncating the weaker wavevector dependence in $N_l$, we can implement high resolutions of the important transfer momentum at moderate numerical cost.

To obtain a set of flow equations for the form-factor dependent channels, i.e. $X^{l,l'}(\bm{q})$ with $X\in \{P,C,D\}$, we take the derivative of the back-transformed form-factor dependent vertex in Eq.~\eqref{eq::formfactorexpansion} and insert two form-factor resolved unities into the contributions in Eqs.~\eqref{eq::contribution1}--\eqref{eq::contribution3}. 
The unities are inserted in such a way that the interactions are separated from the loop kernel, yielding
\begin{align}
    \frac{d}{d\Lambda}P^{l,l'}(\bm{q}) &= \sum_{l_1,l_2}  V^{P}(\bm{q})_{l,l_1} \dot{B}(\bm{q})_{l_1,l_2}^{(-)}V^{P}(\bm{q})_{l_2,l'} \,, \label{eq::ffflowP} \\
    \frac{d}{d\Lambda}C^{l,l'}(\bm{q}) &=  \sum_{l_1,l_2}  V^{C}(\bm{q})_{l,l_1} \dot{B}(\bm{q})_{l_1,l_2}^{(+)}V^{C}(\bm{q})_{l_2,l'}\,, \label{eq::ffflowC}\\
    \frac{d}{d\Lambda}D^{l,l'}(\bm{q}) &=  \sum_{l_1,l_2} \bigl[V^{C}(\bm{q})_{l,l_1} \dot{B}(\bm{q})_{l_1,l_2}^{(+)}V^{D}(\bm{q})_{l_2,l'}  \nonumber \\          
    &\quad\quad+ V^{D}(\bm{q})_{l,l_1} \dot{B}(\bm{q})_{l_1,l_2}^{(+)}V^{C}(\bm{q})_{l_2,l'}  \nonumber \\
    &\quad\quad-2 V^{D}(\bm{q})_{l,l_1} \dot{B}(\bm{q})_{l_1,l_2}^{(+)}V^{D}(\bm{q})_{l_2,l'}  \bigr]\,,\label{eq::ffflowD}
\end{align}
with the form-factor dependent particle-particle ($-$) and particle-hole ($+$) bubble integrals
\begin{align}
        \dot{B}(\bm{q})_{l,l'}^{(\pm)} &=- \int_{p}\frac{d}{d\Lambda}  [G_0^{\Lambda}( i\omega,\bm{q}+\bm{p})\nonumber\\
        &\quad\quad\quad\quad\times G_0^{\Lambda}(\pm i\omega,\pm \bm{q}) ] f_l(\bm{p})f^{*}_{l'}(\bm{p})\,, 
        \label{eq::bubbleform}
\end{align}
cf. App.~\ref{sec:regulator}, and the cross-channel projections
\begin{align}
V^{P}_{l,l'}(\bm{q})&\!=\!\! \int_{\bm{k},\bm{k}'}\!\! f_l(\bm{k})f_{l'}^{*}(\bm{k}') V^{\Lambda}(\bm{k}\!+\! \bm{q},\!-\bm{k}, \bm{k}'\!+\!\bm{q},\!-\bm{k}'),\label{eq::crossprojection0}\\
V^{C}_{l,l'}(\bm{q})&\!=\!\!\int_{\bm{k},\bm{k}'}\!\!  f_l(\bm{k})f_{l'}^{*}(\bm{k}') V^{\Lambda}(\bm{k} + \bm{q}, \bm{k}', \bm{k}' + \bm{q},\bm{k})\,,\\
V^{D}_{l,l'}(\bm{q})&\!=\!\! \int_{\bm{k},\bm{k}'}\!\!  f_l(\bm{k})f_{l'}^{*}(\bm{k}') V^{\Lambda}(\bm{k} + \bm{q}, \bm{k}', \bm{k} ,\bm{k}'+\bm{q})\,,                \label{eq::crossprojection}
\end{align}
where the area of the Brillouin zone is absorbed in the momentum integral: $\int_{\bm{k}}=A_{\mathrm{BZ}}^{-1}\int d\bm{k}$.

\begin{figure}[t!]
\includegraphics[width=\columnwidth]{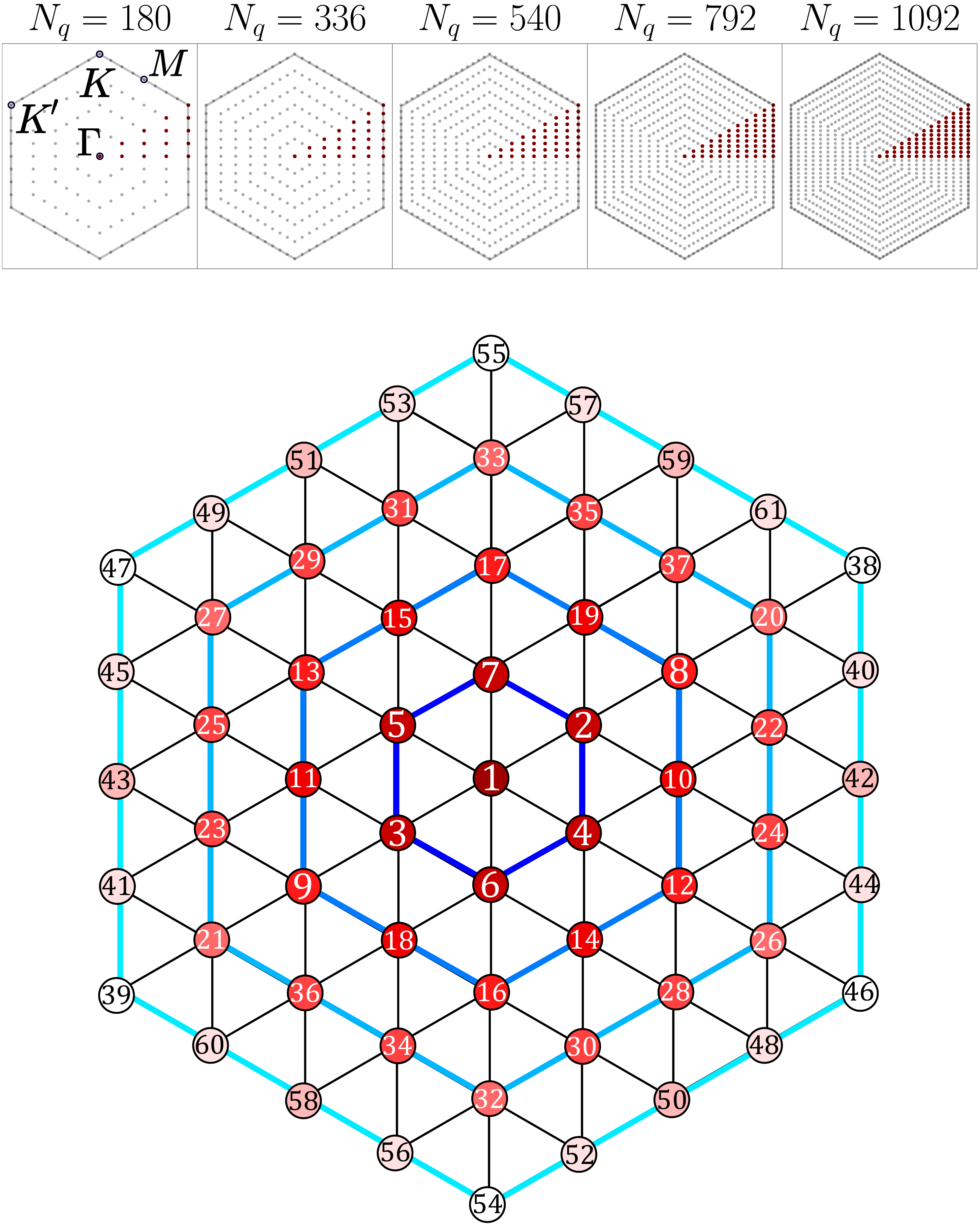}
\caption{\textbf{Momentum mesh and form-factor geometry.} 
Top: Momentum resolution $N_{\bm{q}}$ of the BZ in the transfer momenta. Only a subset (red) of the momentum mesh needs to be calculated. The remaining ones are obtained from symmetries.
Bottom: Numbering of form factors. The shades of red indicate the distance to the origin. The site with number 1 is the onsite form factor, 2--7 are the nearest-neighbor form factors, 10,11,14,15,18,19 are second-nearest-neighbor form factors and  8,9,12,13,16,17 are third-nearest-neighbor form factors. 
The form factor shells are defined as all the form factors sitting on the $s$-th blue hexagon.}
\label{fig::lattices}
\end{figure}

For our implementation we use the momentum mesh shown in Fig.~\ref{fig::lattices} for $N_{\bm q}$ and plane-wave form factors  $f_l(\bm{k})=\exp(i\bm{k}\bm{R}_l)$, which reduce the numerical cost of the projections~\eqref{eq::crossprojection0}--\eqref{eq::crossprojection}.
They also facilitate an intuitive interpretation of form-factor effects on the flow equations in terms of distances in real space.
We group the $f_l$ into shells defined by hexagons of increasing size around a central site, cf. Fig.~\ref{fig::lattices}.
The shells are numbered by  $N_s \in \mathbb{N}_0$ and we include up to $N_s=4$, see App.~\ref{sec::formfactors} for further details. 
In the next section we perform convergence checks in the expansion parameters $(N_{\bm q},N_l)$ and show that a manageable number of transfer momenta and form factors faithfully captures the relevant physics.
The initialization of the RG flow is discussed in App.~\ref{sec:initial}.

\subsection{Analysis of pairing gaps}

In the cases, where the FRG flow signals a pairing instability, we further analyze the type of superconducting pairing. 
To this end, we reconstruct the full superconducting pairing vertex from all form-factor contributions to the $P$ channel.
We use the initial definition of the form factor expansion of the vertices, cf. Eq.~\eqref{eq::formfactorexpansion}, explicitly reading
\begin{align}\label{eq:pairgap}
     \Phi^{P}(\bm{q},\bm{k},\bm{k}') = \sum_{l,l'} P^{l,l'}(\bm{q}) f_l(\bm{k}) f_{l'}^{*}(\bm{k}').
\end{align}
While the divergence of the $P$ channel can occur in different form factor sectors $(l,l')$ at once, the sharp peak is always located $\Gamma$~point of the BZ, i.e. $\bm{q}=0$. 
Therefore, we can reconstruct the superconducting pairing vertex by considering the $\bm{q}=0$ contribution
\begin{align}
\Phi^{P}(\bm{q}=0,\bm{k},\bm{k}'):=\Phi^{P}(\bm{k},\bm{k}')\,,
\end{align}
and using the channel decomposition, Eq.~\eqref{eq::decomposition}.

We can then derive the superconducting interaction from the definition of the effective action.
This can now be treated employing a standard mean-field decoupling within generalized BCS theory~\cite{RevModPhys.63.239}. 
Close to the critical temperature  $T_c$ for the superconducting transition, the gap becomes small, allowing for linearization of the gap equation, i.e.
\begin{align}
    \Delta(\bm{k}) = -\sum_{\bm{k}'} \Phi^{P}(\bm{k},\bm{k}') \frac{\Delta(\bm{k}')}{2\xi_{\bm{k}'}} \mathrm{tanh}\left(\frac{\xi_{\bm{k}'}}{2T_c} \right)\,,
\end{align}
which represents an eigenvalue equation for $\Delta(\bm{k})$. We can approximate its solution by diagonalizing $-\Phi^{P}(\bm{k},\bm{k}')$, which is a $N_{\bm{q}} \times N_{\bm{q}}$ matrix in our TUFRG implementation. 
The eigenvector corresponding to the largest eigenvalue of $-\Phi^{P}(\bm{k},\bm{k}')$ has the highest $T_c$ and therefore determines the structure of the superconducting pairing gap~\cite{RevModPhys.63.239}.

\section{Extended Hubbard model}\label{sec:results}

In this section, we investigate the paradigmatic version of the triangular-lattice model that only includes nearest-neighbor hopping $t=t_1=1$ and all other hopping amplitudes are set to zero, in particular, $t_2=t_3=0$.
For the interactions, we take into account a local or onsite Hubbard interaction $U$ and 
we additionally consider the effect of including a sizable nearest-neighbor interaction $V_1$.
We study the Fermi-surface instabilities that occur in the RG flow near van Hove filling, which for this choice of hopping amplitudes is found at $\mu=2t$. 
Furthermore, we establish convergence of the implementation with respect to the expansion in form-factor shells and momentum resolution.

\subsection{Pure Hubbard model limit}

To draw a connection to previous (FRG) work, we first analyze the case of a pure Hubbard repulsion with an intermediate value of $U=4t$ and we set all non-local interaction contributions to zero, i.e. $V_1=V_2=V_3=0$.
We explore a range of chemical potentials $\mu \in [1.9,2.1]$ around Van Hove filling.
This case was previously studied in Ref.~\cite{PhysRevB.68.104510} using the FRG patching scheme.
The RG flow is initialized at the UV scale with the temperature $T_{\mathrm{UV}}=W$ as the flow parameter, where $W=9t$ is the band width. We then track how the individual channels ($C$, $D$, and $P$) evolve as the temperature scale is lowered. This includes the evolution of their momentum dependence and their overall magnitude.

\subsubsection{Van Hove filling $\mu=2t$}

\begin{figure}[t!]
\includegraphics[width=\columnwidth]{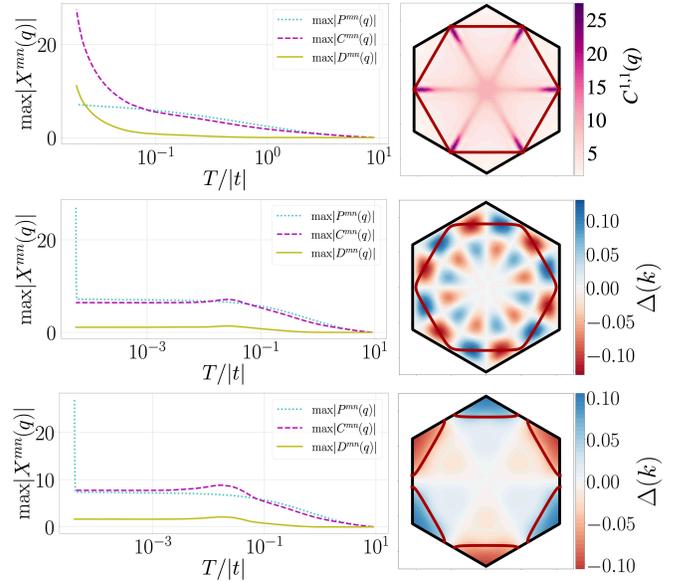}
\caption{\textbf{RG flow and singular vertex for $U\!=\!4t, V_1\!=\!0$} and resolution $N_q=540, N_s=4$. 
Left: Scale-dependence of the $P,C,D$ channel maxima for $\mu=2t$ (top), $\mu=1.96t$ (middle) and $\mu=2.04t$ (bottom). At Van Hove filling $\mu=2t$, the strongest divergence occurs in the magnetic channel $C$. Near Van Hove filling $\mu=1.96t$ or $\mu=2.04t$, the strongest divergence occurs in the pairing channel $P$. 
Right: Momentum dependence of the singular $C$ channel near the instability scale $T_c$ (top), and momentum dependence of the pairing gap constructed from $\Phi^P$, cf. Eq.~\eqref{eq:pairgap} (middle, bottom). The pairing gaps can be classified via the lattice harmonics shown in Fig.~\ref{fig:harmony}. The thick red lines mark the non-interacting Fermi surfaces in each case.}
\label{fig::VHS}
\end{figure}

Right at van Hove filling, the Fermi surface features perfect nesting, which supports relevance of the magnetic channel, i.e. the $C$ channel. The maximal absolute value of the  channels $C$, $D$, and $P$ is shown in Fig.~\ref{fig::VHS}.
We observe a flow to strong coupling at a finite temperature/RG scale of $T^\ast \sim \mathcal{O}(10^{-2}t)$, which is most pronounced in the magnetic $C$ channel. 
The momentum resolution of the C channel at $T^*$ shows that the strongest scattering vectors are located at the $M_i$ points of the BZ.
This signals an instability towards spin-density-wave order with wave vector $M_i$ in agreement with Ref.~\cite{PhysRevB.68.104510}. 
To decide which combination of the three wave vectors $M_i$ is realized in the ordered state, a calculation beyond our current truncation is needed. 
Two candidates, a uniaxial and a chiral SDW state, were identified in previous studies~\cite{PhysRevLett.108.227204,Li_2012,PhysRevB.85.035414,PhysRevLett.101.156402}.

\subsubsection{Below Van Hove filling $\mu<2t$}

For a chemical potential slightly below the Van Hove singularity, e.g., for $\mu=1.96t$, perfect nesting is lost.
Nevertheless, fluctuations in the magnetic channel initially grow due to approximate nesting, but they do not get singular, i.e. the the maximal value of the $C$ channel reaches a maximum value at an RG scale which is close to its instability scale at Van Hove filling.
Still, the RG evolution modifies the effective interaction vertex as the fluctuations in the different channels introduce additional momentum dependencies as compared to the initial interaction. 
This leads to attractive components in the $P$ channel and eventually, at much lower temperatures $T^\ast/t\sim 10^{-5}$, it sharply diverges, signaling an instability towards superconducting order, see Fig.~\ref{fig::VHS}.
We also show the gap function that we obtained from diagonalizing the pairing vertex in Fig.~\ref{fig::VHS}, together with the Fermi line to clearly exhibit the positions of the gap functions' zero-crossings.
The gap function features twelve zero crossings along the Fermi surface and corresponds to spin-singlet pairing in the $A_2$ irrep (cf. Fig.~\ref{fig:harmony}), which can also be referred to as $i$-wave superconducting instability.
This type of instability for the present model parameters was previously discussed in related FRG studies~\cite{PhysRevB.68.104510,scherer2021mathcal}.

\subsubsection{Above Van Hove filling $\mu>2t$}

A chemical potential slightly above the Van Hove singularity, e.g., for $\mu=2.04t$, also makes the nesting of the Fermi surface only approximate.
Similarly to the case $\mu<2t$, this allows a pairing instability to develop and we find that at low temperatures $T^\ast/t \sim 10^{-4}$, the $P$ channel sharply diverges again.
However, the symmetry of the leading pairing instability is different for the low and high doping sides if $V_1=0$.
For $\mu=2.04t$, the instability is towards superconductivity with a spin-triplet pairing function in the $B_1$ irrep (cf. Fig.~\ref{fig:harmony}), i.e. an $f$-wave superconductor, see Fig.~\ref{fig::VHS}. 
The Fermi surface consists of pockets around the $K,K'$ points so that an f-wave gap does not have any nodes on the Fermi surface.
This state was previously described in Ref.~\cite{PhysRevB.98.174515}.

\subsection{Inclusion of nearest-neighbor interaction}

\begin{figure}[t!]
   \includegraphics[width=\columnwidth]{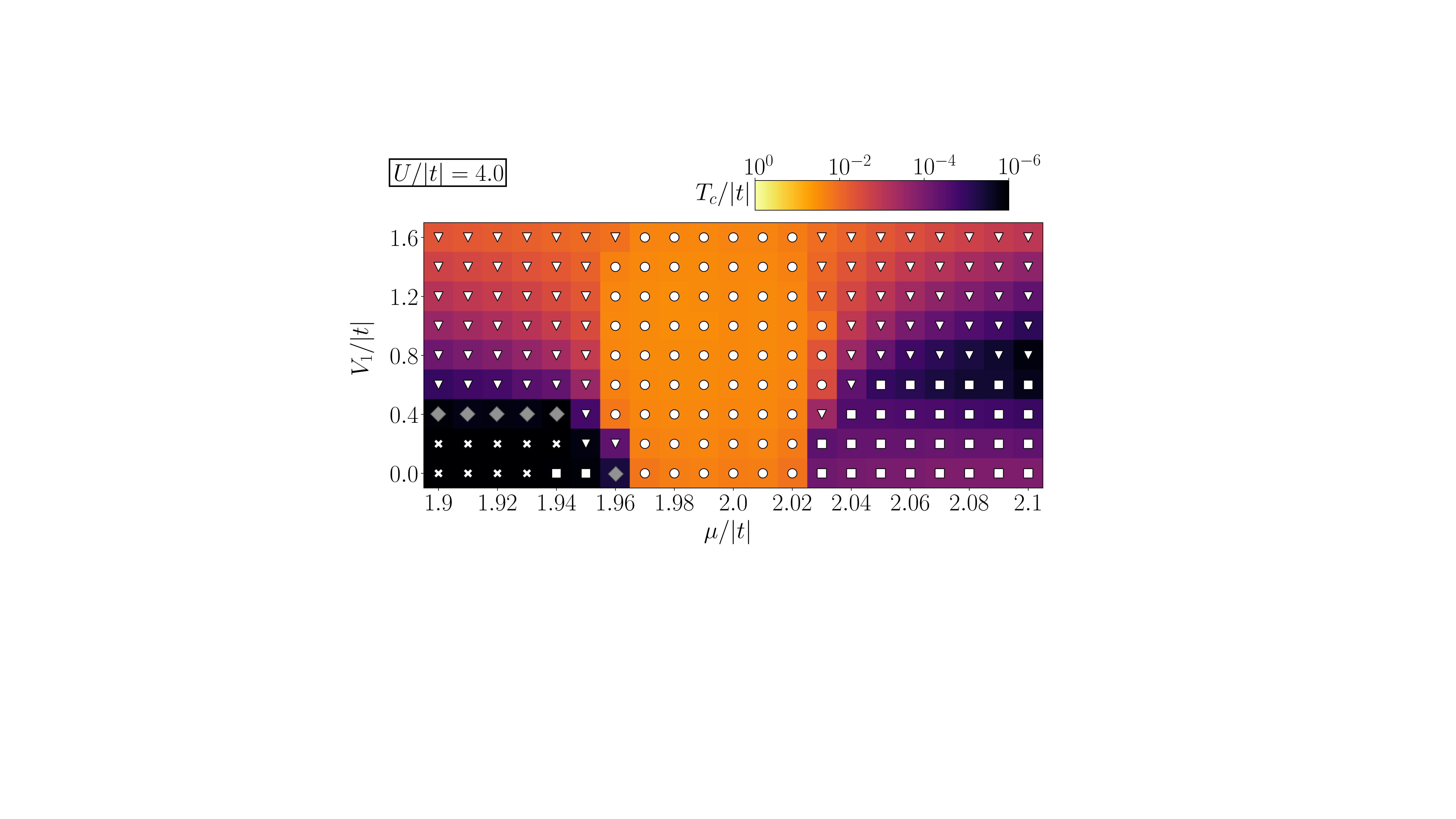}
    \caption{\textbf{Paradigmatic phase diagram} of the triangular-lattice extended Hubbard model for $U=4t$ near $3/4$ filling exhibiting SDW ($\circ$), 
    $g$SC ($\triangledown$), $i$SC ($\diamond$), and $f$SC ($\square$) phases.
    Parameters, where no instability occurs for $T/t \geq 10^{-6}$ are marked as metallic ($\times$).
    The symbols for the $i$SC are shown in gray shading to indicate  fragility with respect to the expansion in form-factor shells $N_s$, cf. Sec.~\ref{subsec:convergence}.}
    \label{fig:PhaseDiagram1}
\end{figure}

To study the effect of the nearest-neighbor interaction on the Van Hove scenario, we investigate the parameter range $0\leq V_1/t\leq 1.6$ around Van Hove filling for $1.9\leq \mu/t\leq 2.04$.
For larger $V_1/|t|$ we find an instability towards a charge-density-wave state.
We show the resulting phase diagram of ordering tendencies in Fig.~\ref{fig:PhaseDiagram1}. 
We find that the SDW is unimpressed by the presence of even a sizable $V_1$; 
it slightly extends in $\mu$ and its critical temperature hardly changes. 
In contrast, the pairing instability is strongly affected when $V_1$ is included and a rich phase diagram unfolds.
Increasing $V_1$ extends the range of chemical potentials, where a superconducting instability is found substantially and also provides regions in parameter space with significantly enhanced critical temperatures.

\begin{figure}[t!]
\includegraphics[width=\columnwidth]{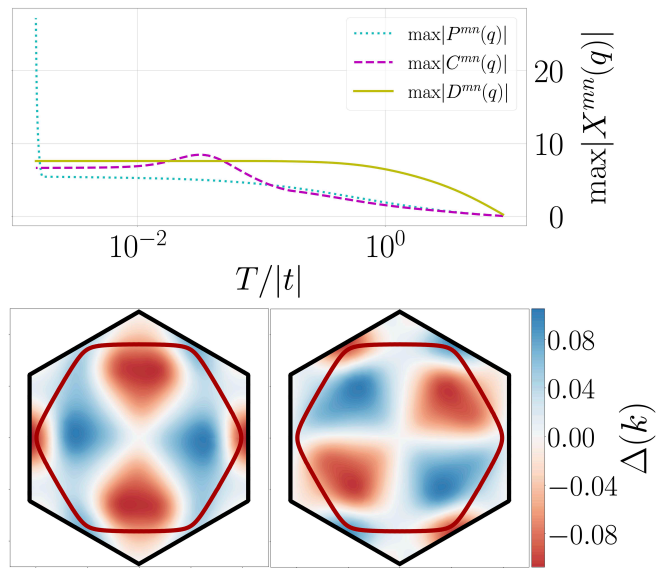}
\caption{\textbf{RG flow and momentum dependence of 
pairing solutions} for $\mu=1.92t, U=4.0t, V_1=1.2$ (momentum resolution $N_{\bm{q}}=540$, and form factor shells $N_s=4$).
Top: Scale-dependence of the channel maxima. In contrast to the case $V_1=0$, the density channel $D$ initially increases the strongest and then saturates at a comparable level as the magnetic channel $C$. At smaller scales the pairing channel $P$ develops an instability. Bottom: The gap equation has two independent solutions, which are dominated by the second-nearest neighbor harmonic of the $E_2$ irrep (cf.~Fig.~\ref{fig:harmony}). The non-interacting Fermi surface is marked by the thick, red line. We find similar pairing solutions for $\mu>2t$ and sizable $V_1$, where the Fermi surface consists of pockets around $K,K'$.}
\label{fig::gwaveSC}
\end{figure}

Furthermore, a non-local contribution of the Coulomb interaction can induce a stronger momentum-dependence of the pairing symmetry because the pair formation is pushed to larger distances to avoid not only on-site but also nearest-neighbor repulsion.
We have previously observed this effect in Ref.~\cite{scherer2021mathcal} and we can corroborate this finding with the present TUFRG approach.
We find that both, the $i$- and $f$-wave pairing states get destroyed by $V_1$.
Instead, for sizable values of the nearest-neighbor interaction $V_1/t \gtrsim 0.5$, we obtain an extended pairing region above and below Van Hove filling which is in the two-dimensional $E_2$ irrep. 
This implies that the gap equation possesses two degenerate pairing solutions with the same critical temperature, cf. Fig.~\ref{fig::gwaveSC}.

Remarkably, these pairing solutions are dominated by the second-nearest neighbor harmonics of the $E_2$ irrep, cf. Fig.~\ref{fig:harmony}, which -- according to the number of nodes along the Fermi line -- is dubbed $g$-wave pairing. 
In comparison, pure nearest-neighbor lattice harmonics in the $E_2$ irrep would be classified as $d$-wave due to their number of nodes. 
In principle, these lattice harmonics can mix because they belong to the same irrep.
The dominance of the $d$- or $g$-wave contributions can be inferred from the number of zero crossings of the pairing solutions, which we have extracted from the FRG data, cf. Fig.~\ref{fig:harmony}  .
Both filling cases clearly exhibit eight nodes along the Fermi surfaces~\footnote{Note that the number of nodes of the pairing solutions does not simply translate to the number of nodes of the superconducting gap in the case of 2D irreps, because the superconducting gap is formed by a specific linear combination of the two pairing solutions. Depending on the linear combination, the superconducting gap can be gapless or possess nodes on the Fermi surface.}, i.e. the FRG data supports the formation of $g$-wave pairing above and below the Van-Hove point in the investigated range of fillings.
This means that the largest contribution comes from the second-nearest neighbor harmonics in agreement with the expectation that $V_1$ pushes the pair formation outwards.

\subsection{Convergence}\label{subsec:convergence}

\begin{figure}[t!]
\includegraphics[width=\columnwidth]{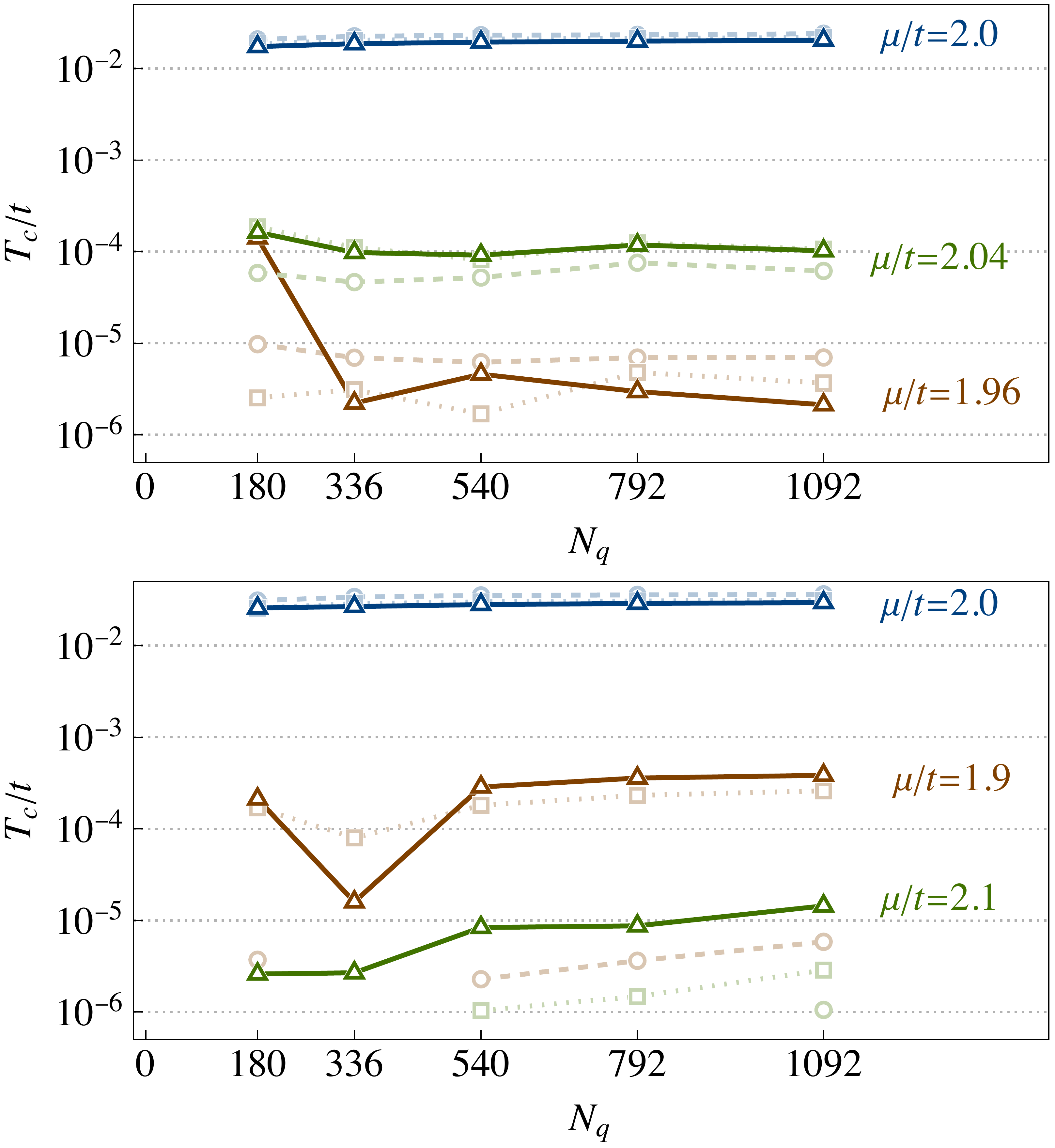}
\caption{\textbf{Resolution dependence of $T_c$} for $U\!=\!4t, V_1\!=\!0$ (top) and $U\!=\!4t, V_1=t$ (bottom).
Convergence of critical temperature as a function of momentum resolution $N_{\bm{q}}$ and number of form factor shells for $N_s=4$ (solid lines), $N_s=3$ (dotted lines), and $N_s=2$ (dashed lines).}
\label{fig::convergenceN20}
\end{figure}

To establish convergence of our numerical results, we explore their dependence on the momentum resolution $N_{\bm{q}}$ and form factor expansion $N_s$.
Explicitly, we use $N_s=2,3,4$, which translates into corresponding number of form factors in the following way (see Fig.~\ref{fig::lattices}):

\begin{table}[h!]
\centering
\begin{tabular}{c|c|c}
 \hline\hline
 $N_s$  & $N_l$ & \shortstack{Corresponding nearest-neighbor shells} \\
 \hline
$2$ & 19 & 0th - 3rd nearest-neighbors \\
 $3$ & 37 & 0th - 5th nearest-neighbors \\
 $4$ & 61 & 0th - 8th nearest-neighbors \\
 \hline\hline
\end{tabular}
\end{table}
\noindent To address the convergence with respect to $N_{\bm{q}}$, we run the TUFRG flow and extract the critical temperature $T_c/t$ at fixed $U,V_1$ and three choices of the chemical potential below, at, and above Van Hove filling for $N_{\bm{q}}\in\{180,336,540,792,1092\}$. We consider two cases: $V_1=0$ with $\mu/t=1.96, 2.0, 2.04$ and $V_1=t$ with $\mu/t=1.9, 2.0, 2.1$, cf. Fig.~\ref{fig::convergenceN20}. 
We conclude that a momentum resolution of $N_q=540$ is sufficiently good and higher resolutions appear to only quantitatively change the results for the critical temperature.

\begin{figure}[t!]
\centering
    \includegraphics[width=\columnwidth]{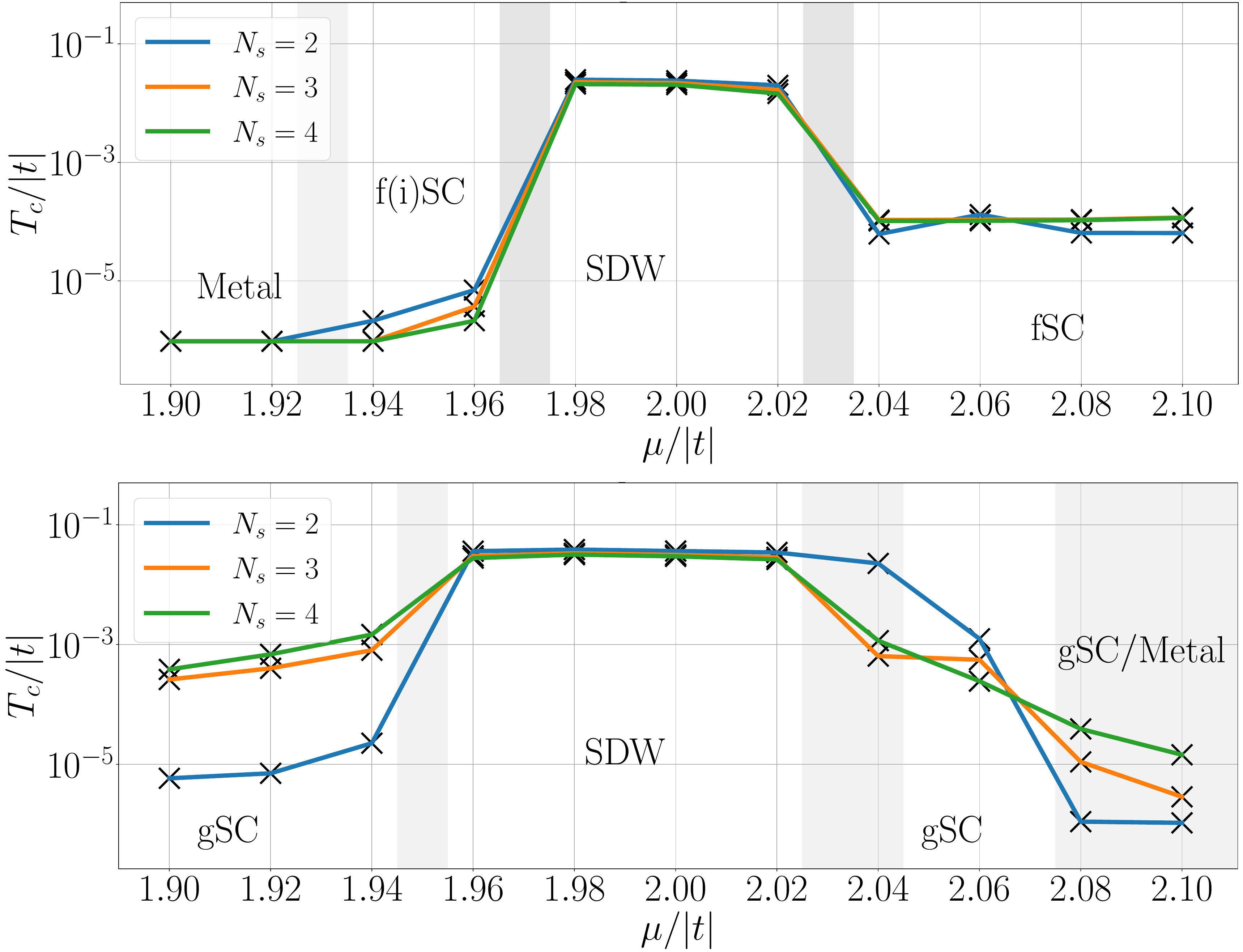}
    \caption{\textbf{Phase diagram and convergence checks} for $U=4t_1$ and $V_1=0$ (top) or $V_1=t$ (bottom) with $N_q=1092$.
    Instability types are labeled and phase boundaries are indicated by gray shadings.
    Quantitative convergence of the results as a function of $N_s$ is generally very good.
    Exceptions can be found in the transition regimes of different types of instabilities, i.e. near $\mu/t \sim 2.04$, where for smaller $N_s=2$ the SDW is the leading instability and for larger $N_s$ we find $g$-wave pairing. 
    For $V_1=0$ and $\mu \lesssim 1.96$ we need to take $N_s=4$ to resolve $i$-wave pairing. 
    For $V_1=t$, the type of superconducting instability does not change for $N_s\geq 2$. We show an equivalent figure for $N_q=540$ in App.~\ref{app::shells}.}
    \label{fig::convergence1092}   
\end{figure}

To compare the effect of including different numbers of shells, we choose a high resolution $N_{\bm{q}}=1092$ and vary $N_s$.
For $U/t=4, \mu/t \in [1.9,2.1]$, and $V_1/t\in\{0,1\}$, the dependence of the emerging instability on $(N_s, N_{\bm{q}})$ is shown in Figs.~\ref{fig::convergence1092}. 
We find that the number of form factor shells can be important to determine the instability type at phase boundaries and to faithfully resolve the pairing symmetry.
For example, for $N_{\bm{q}}=1092$ at $\mu=2.04t$ in Fig.~\ref{fig::convergence1092}, smaller $N_s=2$ yields an SDW instability with higher $T_c$, while a better resolution $N_s=3,4$ gives rise to a pairing instability.
In such cases the phase boundary has to be determined with care.
Regarding the pairing symmetry, in particular, the $i$-wave superconducting instability requires to take into account $N_s=4$ shells. 
If not enough form-factor shells are taken into account, the leading superconducting instability is found to be in the two-dimensional $E_1$ irrep ($p$-wave) for $N_s=2$, or in the $B_2$ irrep ($f$-wave) for $N_s=3$. 
For the other instabilities which we reported for the pure Hubbard model, we found that already a lower expansion with $N_s=2$ is sufficient. For the $g$-wave instability that occurs for larger values of $V_1$, we find that qualitatively the second-nearest-neighbor harmonics are obtained with $N_s\geq2$. 

Generally, we find reasonable convergence in $(N_s, N_{\bm{q}})$ and apart from the above mentioned exceptions, the choice $(N_s, N_{\bm{q}})=(2,540)$ delivers reliable results.
To keep the study as unbiased as possible, all results shown in the phase diagrams (Fig.~\ref{fig:PhaseDiagram1} and \ref{fig:PhaseDiagram2}) are calculated with $N_s=4$. This significantly increases the number of form-factors which have been used for hexagonal lattices compared to previous works.

\section{Moir\'e TMD heterobilayers}\label{sec:results2}

Moir\'e quantum materials have been  established as a solid platform to study strongly-correlated systems in a controlled and tunable way.
Specifically, hetero-bilayers of TMDs at small angle have been shown to realize the extended triangular-lattice Hubbard model where the whole range of band fillings is experimentally accessible and interactions can be tuned within a large window. 
As a concrete example we study the AA-stacked hetero-bilayer of WSe${}_2$/MoS${}_2$ at vanishing twist angle.
Here, the highest spin-polarized valence band from 
WSe${}_2$ contributes an isolated flat band at the Fermi level to the moir\'e band structure~\cite{PhysRevLett.121.026402}.
This isolated band can be accurately described by a tight-binding model with up to third-nearest-neighbor hopping, $t_1\approx -2.5\,\mathrm{meV}, t_2\approx 0.5\,\mathrm{meV}, t_3\approx 0.25\,\mathrm{meV}$, cf. Sec.~\ref{sec:model}.
The resulting band structure features a Van Hove singularity near $-5.5\,\mathrm{meV}$ and the Fermi surface at that filling is approximately nested. 
The interactions can be tuned in strength and range using dielectric environments or screening layers.
Explicitly, the system can be tuned into the intermediate interaction regime, which we associate roughly with an onsite repulsion $U/t\approx 4$.
To explore the effect of tunability in the range of the interactions, we vary $V_1$ again but additionally include the longer-ranged interactions $V_2$ and $V_3$, choosing fixed ratios with $V_1$, i.e. $V_2/V_1\approx 0.357$ and $V_3/V_1\approx 0.260$ as estimated in Ref.~\cite{zhou2021quantum}.

\begin{figure}[t!]
   \includegraphics[width=\columnwidth]{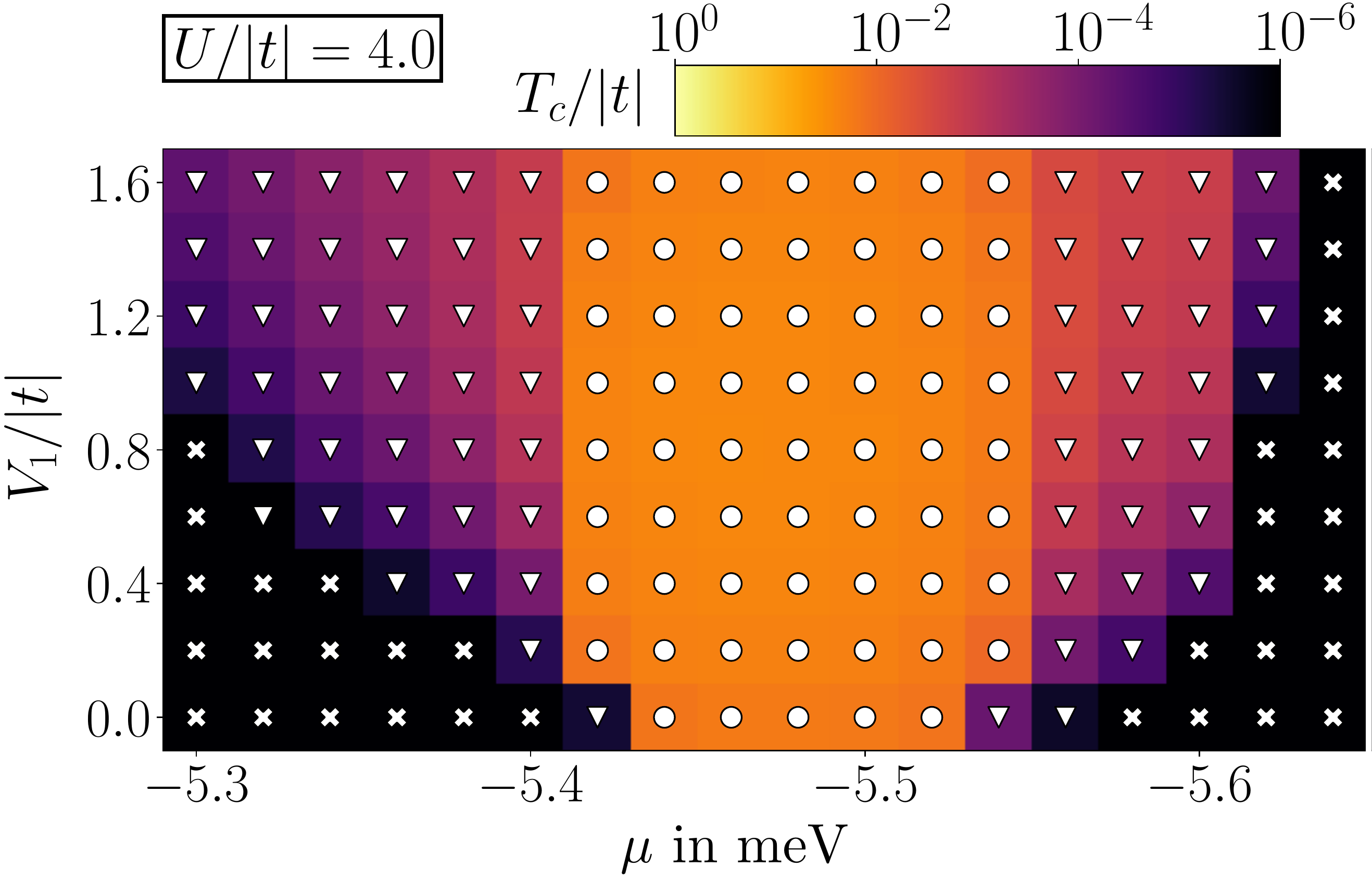}
    \caption{\textbf{Moir\'e TMD phase diagram} of the effective extended triangular-lattice Hubbard model describing TMD hetero-bilayers near $1/4$ filling. The phase diagram includes VDW ($\circ$), $g$SC ($\triangledown$), and metallic ($\times$) phases.}
    \label{fig:PhaseDiagram2}
\end{figure}

In Fig.~\ref{fig:PhaseDiagram2}, we present the correlated phase diagram 
that contains the instabilities we predict using the TUFRG as function of $\mu$ and $V_1$. It is qualitatively very similar to the paradigmatic case in Sec.~\ref{sec:results}; directly around Van Hove filling, we obtain a SDW instability bounded by pairing instabilities. However, the additional hoppings and interactions decrease the critical temperature and even completely suppress some pairing instabilities at smaller $V_1$. We now only find pairing states which belong to the irrep $E_2$ with the largest contribution coming from second-nearest-neighbor harmonics, i.e. we can classify them as $g$-wave based on their number of nodes along the Fermi surface (as in Fig.~\ref{fig::gwaveSC}). For $\mu>-5.5$meV, where the Fermi surface is closed around $\Gamma$, this confirms the previous low-resolution FRG calculation~\cite{scherer2021mathcal} with the exception that we do not find the $i$-wave pairing instability at $V_i=0$ that was already marked as fragile in Ref.~\onlinecite{scherer2021mathcal}.  Instead, the $i$-wave pairing is destroyed when more distanced hoppings $t_2,t_3$ are included. 
We extend the phase diagram to $\mu<-5.5$meV, where the Fermi surface consists of pockets around $K,K'$. We find the same $g$-wave pairing instability as for $\mu>-5.5$meV at larger $V_i$, while the $f$-wave pairing for small $V_1$ in Fig.~\ref{fig:PhaseDiagram1} also disappears when $t_2$ and $t_3$ are included.

We emphasize that the $g$-wave pairing is a robust result that we obtain with both models. This stronger momentum dependence arises due to the inclusion of a sizable nearest-neighbor interaction $V_1>0$. It is based on pairing between second-nearest neighbors as the classification in terms of lattice harmonics shows so that on-site and nearest-neighbor repulsion can be avoided.

\section{Topological superconductivity}\label{sec:g+ig}

If a superconducting state develops out of the $g$-wave pairing instability, a specific linear combination of the two pairing solutions is formed. 
It was argued that a chiral linear combination $g\pm i g$ 
minimizes the Landau free energy and thus makes up the ground state\cite{scherer2021mathcal}. 
Such a $g\pm i g$ state fully gaps the Fermi surface despite the high number of nodes in the single pairing solutions.
It also breaks time-reversal symmetry and is topologically non-trivial with a non-zero winding number which can be defined via
\begin{align}
 \mathcal{N} = \frac{1}{4\pi}\int_{\mathrm{BZ}}d^2k\, \vec{m}\cdot\left(\frac{\partial \bm{m}}{\partial k_x}\times\frac{\partial \bm{m}}{\partial k_y}\right)\,,
\end{align}
with the pseudo-spin vector
\begin{align}
\vec{m}=\frac{1}{E_{\vec k}}(\mathrm{Re}\Delta_{\bm{k}},\mathrm{Im}\Delta_{\bm{k}},\xi_{\bm{k}})^T\,.
\end{align}
We find that in the whole parameter range, where the pairing gap in the $E_2$ irrep occurs, the second-nearest neighbor harmonics dominate.
The implied $g+ig$ state results in a Chern number of $|\mathcal{N}|=4$.
This is in contrast to $|\mathcal{N}|=2$ for a $d+id$ state, which would be a possible ground state if the nearest-neighbor harmonics dominated.
The Chern number is directly proportional to the quantum spin and thermal Hall conductance in such a topologically non-trivial superconducting state.
Therefore, the higher-harmonic gap function can manifest itself experimentally by enhanced quantum spin and thermal Hall responses~\cite{scherer2021mathcal}.

\section{Conclusion}\label{sec:conclusion}

In this work, we have taken recent developments in the field of correlated moir\'e materials as motivation to revisit the phase diagram of correlated electrons on the triangular lattice with extended Hubbard interactions near Van Hove filling.
To that end, we implemented and carefully benchmarked the TUFRG scheme in a numerically efficient way.
We studied two versions of the extended Hubbard model: a minimal, paradigmatic one with only $t_1,U$ and $V_1$ non-zero and a realistic one with parameters for the moir\'e hetero-bilayer WSe${}_2$/MoS${}_2$.
In both cases, we found strong evidence that in an intermediate-interaction regime, sizable nearest-neighbor interactions induce a chiral superconducting $g+ig$ instability, which is realized in higher lattice harmonics indicating pairing between second-nearest-neighbor sites.
The $g+ig$ state corresponds to a fully gapped, topological superconductor and features enhanced quantum Hall responses as compared to the chiral $d+id$ superconducting state made of the 
nearest-neighbor harmonics.

To obtain our results, we capitalized on the advantages provided by the TUFRG framework, i.e. inclusion of all types of correlations on equal footing within a (realistic) microscopic lattice model and high momentum resolution of the BZ allowing for a detailed analysis of the emergent instabilities.
Both are crucial in the present context. 
We need the unbiased approach that includes all correlations to analyze the interplay between competing orders around Van Hove filling and, in particular, to describe electronic pairing from repulsive bare interactions. 
A high momentum resolution is necessary to correctly resolve the pairing symmetries and their leading contributions from different lattice harmonics.
We hope to inspire further studies to challenge our results with other quantum many-body methods, possibly facilitated by our analysis of the paradigmatic version of the model.

In the future, it will be interesting to extend our present TUFRG implementation to systems without an SU(2) invariance in the (pseudo-)spin or valley degrees of freedom.
Applications of this extension include moir\'e TMD homobilayers, or unconventional superconductors with spin-orbit couplings.
Similar efforts are currently undertaken by other groups~\cite{beyer2022reference}.

\begin{acknowledgments}
We thank Ammon Fischer, Ravn Henkel, Dante Kennes, Dominik Kiese, Lennart Klebl, Thomas Sch\"afer and Simon Trebst for helpful discussions. 
MMS acknowledges support by the Deutsche  Forschungsgemeinschaft (DFG, German Research Foundation) through the DFG Heisenberg programme (project id 452976698) and SFB 1238 (project C02, project id 277146847).
\end{acknowledgments}

\appendix

\section{Regulator}\label{sec:regulator}

Generally, in the FRG approach the regularization can be implemented in various ways, e.g., employing a momentum cutoff, a frequency cutoff, or a temperature cutoff scheme and the concrete choice should be guided by the problem at hand\cite{RevModPhys.84.299}.
In the present work, we choose the temperature cutoff scheme~\cite{PhysRevB.64.184516}.
To that end, the fermionic fields are redefined such that only the quadratic part of the action is temperature dependent. 
This leads to a temperature-scaled free propagator and the scale derivative is turned into a temperature derivative, i.e.
\begin{align}
G_0(i\omega,\bm{k}) &\rightarrow  G_0^T(i\omega,\bm{k})= \frac{T^{1/2}}{i\omega-\xi(\bm{k})}\,,\\
\frac{d}{d\Lambda} &\rightarrow  \frac{d}{dT}\,.
\end{align}
Inserting these expressions into the form-factor-dependent bubbles in Eq.~\eqref{eq::bubbleform} and carrying out the Matsubara summation yields the bubble integrals
\begin{align}
 \dot{B}(\bm{q})_{l,l'}^{(+)}\! &=\!+\!\!\int_{\bm{p}}\! \frac{n_F'(\xi(\bm{q}\!+\!\bm{p})\!)\!-\!n_F'(\xi(\bm{q})\!) }{ \xi(\bm{q}\!+\!\bm{p})\!-\! \xi(\bm{q})}f_l(\bm{p})f_{l'}^{*}(\bm{p}),         \label{eq::bubble1} \\[8pt]
 \dot{B}(\bm{q})_{l,l'}^{(-)}\! &=\!-\!\!\int_{\bm{p}}\!\! \frac{n_F'(\xi(\!\bm{q}\!+\!\bm{p})\!)\!+\!n_F'(\xi(\!-\bm{q})\!)}{ \xi(\bm{q}\!+\!\bm{p})\!+\! \xi(-\bm{q})}f_l(\bm{p})f_{l'}^{*}(\bm{p}),         \label{eq::bubble2}
\end{align}
where $n_F'(x)$ denotes the temperature derivative of the Fermi function, i.e. $n_F'(x) = \frac{d}{dT} n_F(x)$.

\section{Form factors}\label{sec::formfactors}

To solve the RG flow Eqs.~\eqref{eq::ffflowP}--\eqref{eq::ffflowD} numerically, we need to make a specific choice for the form factors, cf. Eq.~\eqref{eq::formfactorexpansion}.
In this work, we employ an expansion in terms of plane waves, i.e. $f_l(k)=\exp\left(i\bm{k}\bm{R}_l\right)$ where $\bm{R}_l$ is a Bravais lattice vector of the triangular lattice and this choice naturally fulfills the conditions in Eqs.~\eqref{eq::unity1} and~\eqref{eq::unity}.
The plane-wave expansion has three advantages, which we will explain in the following: 
(1)~it is possible to apply a physical interpretation of the included form factors, 
(2)~it becomes apparent which truncation of form factors is reasonable, and
(3)~it simplifies the cross-channel projections, cf. Eqs.~\eqref{eq::crossprojection0}--\eqref{eq::crossprojection}.

To expose the physical interpretation of this choice of form factors, we consider the interaction vertex, cf. Eq.~\eqref{eq::Gamma4}, and insert the channel decomposition Eq.~\eqref{eq::decomposition}. 
Four terms emerge and we focus the discussion on the $P$ channel. 
Relabeling the wave-vectors with the strong momentum $\bm{q}$ and the weak momenta $\bm{k},\bm{k}'$ and exploiting momentum conservation, the superconducting interaction reads
\begin{align}
\Gamma^{P}=& \frac{1}{2}\int_{\bm{q},\bm{k},\bm{k}'}\Phi^{P}(\bm{q},\bm{k},\bm{k}')    \nonumber\\
&\times \sum_{\sigma,\sigma'} \bar{\psi}_\sigma(\bm{q}+\bm{k}) \bar{\psi}_{\sigma'}(\bm{k}) \psi_{\sigma'}(-\bm{k}') \psi_{\sigma}(\bm{q}+\bm{k}')\,.
\label{eq::SC}
\end{align}
This structure can be identified on the l.h.s. of Fig.~(\ref{fig:decompo}).
Expanding $\Phi^{P}(\bm{q},\bm{k},\bm{k}')$ as in Eq.~\eqref{eq::formfactorexpansion} yields
\begin{align}
\Gamma^{P} &= \frac{1}{2}\int_{\bm{q}} \sum_{\substack{l,l\\ \sigma,\sigma'}}P^{l,l'}(\bm{q}) \underbrace{\left(\int_k \bar{\psi}_\sigma(\bm{q}+\bm{k}) \bar{\psi}_{\sigma'}(-\bm{k}) f_l(\bm{k})\right)}_{\bar{F}(\bm{q})_l^{\sigma,\sigma'}} \nonumber \\
&\quad\quad\quad\quad \times \underbrace{\left(\int_{\bm{k}'} \psi_{\sigma'}(-\bm{k}') \psi_{\sigma}(\bm{q}+\bm{k}') f^{*}_{l'}(\bm{k}') \right)}_{F(\bm{q})_{l'}^{\sigma,\sigma'}}\,.
\label{eq::SC1}
\end{align}
The fermion bilinears $F(\bm{q})_l^{\sigma,\sigma'}$ decouple from each other.
To expose the effect of the form factors $f_l(k)$ on these bilinears we transform the fields into real space,
\begin{align}\label{eq:bilinears}
F_{l}^{\sigma,\sigma'}(\bm{q}) &= \int_{\bm{k}'} \psi_{\sigma'}(-\bm{k}') \psi_{\sigma}(\bm{q}+\bm{k}') f^{*}_{l'}(\bm{k}') \nonumber \\
              &= \sum_{\bm{R}} \psi_{\sigma'}(\bm{R}-\bm{R}_l) \psi_{\sigma}(\bm{R}) e^{i\bm{q}\bm{R}} \,.
\end{align}
This shows that $F_{l}^{\sigma,\sigma'}(\bm{q})$ includes all two-fermion combinations, which can be connected by the lattice vector $\bm{R}_l$ and therefore the inclusion of form factor $f_l(\bm{k})$ can be interpreted as including all fermionic bilinears with distance $\bm{R}_l$ in the interaction Eq.~\eqref{eq::SC1}.

Most common Fermi-surface instabilities are caused by long-range fluctuations that are resolved via the transfer momentum $\bm q$ of the different channels. 
Fluctuations in the additional $\bm k,\bm k'$ dependence typically correspond to a short spatial range, i.e. they occur within a radius of a couple of unit lattice vectors. 
Therefore, we truncate the form factors with respect to the distance occurring in the bilinears, cf. Eq.~\eqref{eq:bilinears}. 
To that end, we introduce the notion of form-factor shells $N_s$. The $s$-th form factor shell will then include all real space vectors which point to a lattice point lying on the $s$-th hexagon as indicated in Fig.~\ref{fig::lattices}.
We note that there is also a zeroth form-factor shell including only the onsite contribution $\bm{R}_l=0$.

This truncation involving the form-factor shells leads to a numerical advantage of the TUFRG scheme. 
While the unity in Eq.~\eqref{eq::unity} is only exact if all lattice vectors $\bm{R}_l$ are included, we expect the form factors belonging to high form factor shells to have little actual impact. 
Therefore, we choose to include only a few short-range form factors in the vertices,  Eqs.~\eqref{eq::ffflowC}--\eqref{eq::ffflowD}, for the actual calculations. 
The size of the vertices scales with $N_{\bm{q}} \times N_l^2$. 
For high momentum resolution $N_{\bm{q}}$, this scaling becomes superior to the  $N_{\bm{q}}^3$~scaling in patch-RG schemes.

Moreover, the plane waves simplify the cross projections in Eqs.~\eqref{eq::crossprojection0}--\eqref{eq::crossprojection}. 
Employing the decomposition Eq.~\eqref{eq::decomposition} and explicitly using the plane waves, the projections simplify from a double integration over the BZ to a simple sum of the included form factors, i.e.
\begin{align}\label{eq::crossprojectionsexplicit1}
V^{P}_{l,l'}(\bm{q})\! =\! V^{P,0}_{l,l'}(\bm{q})\!+\!V^{P\veryshortarrow C}_{l,l'}(\bm{q})\!+\!V^{P\veryshortarrow D}_{l,l'}(\bm{q})\!+\!P_{l,l'}(\bm{q})\,,
\end{align}
where
\begin{align}
V^{P\veryshortarrow C}_{l,l'}(\bm{q}) &=\! \sum_{L} \!\tilde{C}_{\bm{R}_{L},-\bm{R}_{L}+\bm{R}_{l}+\bm{R}_{l'}} (\!-\!\bm{R}_{L}\!+\!\bm{R}_{l'}) e^{-i(\bm{R}_{L}\!-\!\bm{R}_{l'})\bm{q}}\,,         \nonumber \\ 
V^{P\veryshortarrow D}_{l,l'}(\bm{q}) &=\! \sum_{L}\! \tilde{D}_{\bm{R}_{L},-\bm{R}_{L}+\bm{R}_{l}-\bm{R}_{l'}} (\!-\!\bm{R}_{L}\!-\!\bm{R}_{l'}) e^{-i\bm{R}_{L}\bm{q}}\,.\nonumber
\end{align}
Accordingly, we find
\begin{align}
V^{C}_{l,l'}(\bm{q})\!=\!V^{C,0}_{l,l'}(\bm{q})\!+\!V^{C \veryshortarrow P }_{l,l'}(\bm{q})\!+\!V^{C \veryshortarrow D}_{l,l'}(\bm{q})\!+\! C_{l,l'}(\bm{q})\,,
\end{align}
with
\begin{align}
V^{C\veryshortarrow P}_{l,l'}(\bm{q}) &=\!  \sum_{L}   \tilde{P}_{\bm{R}_{L},-\bm{R}_{L}+\bm{R}_{l}+\bm{R}_{l'}} (\!-\!\bm{R}_{L}\!+\!\bm{R}_{l'}) e^{-i(\bm{R}_{L}\!-\!\bm{R}_{l'})\bm{q}}\,,\nonumber\\
V^{C\veryshortarrow D}_{l,l'}(\bm{q}) &=\! \sum_{L}    \tilde{D}_{\bm{R}_{L},\bm{R}_{L}-\bm{R}_{l}+\bm{R}_{l'}} (-\bm{R}_{l})e^{-i\bm{R}_{L}\bm{q}}\,,\nonumber
\end{align}
and finally
\begin{align}\label{eq::crossprojectionsexplicit3}
V^{D}_{l,l'}(\bm{q})\!=\!V^{D,0}_{l,l'}(\bm{q})\!+\! V^{D \veryshortarrow P}_{l,l'}(\bm{q})\!+ \!V^{D \veryshortarrow C}_{l,l'}(\bm{q})\!+\! D_{l,l'}(\bm{q})\,,
\end{align}
with
\begin{align}
V^{D\veryshortarrow P}_{l,l'}(\bm{q}) &= \sum_{L} \tilde{P}_{\bm{R}_{L},\bm{R}_{L}-\bm{R}_{l}-\bm{R}_{l'}} (-\bm{R}_{l}) e^{-i(\bm{R}_{L} - \bm{R}_{l'})\bm{q}}\,,  \nonumber \\
V^{D\veryshortarrow C}_{l,l'}(\bm{q}) &= \sum_{L} \tilde{C}_{\bm{R}_{L},\bm{R}_{L}-\bm{R}_{l}+\bm{R}_{l'}} (-\bm{R}_{l}) e^{-i\bm{R}_{L}\bm{q}}.
\nonumber
\end{align}
The $V^{X,0}_{l,l'}(\bm{q})$ are determined by the projection of the initial interaction into the channels, cf. Sec.~\ref{sec:initial}. 
The $\tilde{X}_{l,l'}$ are the Fourier-transformed channels, e.g., for $P$
\begin{align}
    \tilde{P}_{l,l'}(R_i) = A^{-1}_{\mathrm{BZ}} \int \ d\bm{p}\, P_{l,l'}(\bm{p}) e^{-i\bm{p}\bm{R}_i} .
\end{align}
The sum $\sum_L$ runs over all included form factors, i.e. for our application this will include $R_1$ to $R_{61}$, see Fig.~\ref{fig::lattices}.

\section{Initial conditions}\label{sec:initial}

The initial conditions for the RG flow of $V^\Lambda$ are determined by the Fourier-space interactions corresponding to the Hamiltonian parameters $U, V_1,V_2$, and $V_3$.
We rewrite the interactions accordingly and use Eq.~\eqref{eq::Gamma4} to extract the expression for $V^{\Lambda,0}(\bm{k}_1,\bm{k}_2,\bm{k}_3,\bm{k}_4)$, which is then plugged into Eqs.~\eqref{eq::crossprojection0}--\eqref{eq::crossprojection} to derive explicit expressions for the  $V^{X,0}_{l,l'}(\bm{q})$.
We find
\begin{align}
V^{P,0}_{\bm{R}_1,\bm{R}_1}(\bm{q})&=V^{C,0}_{\bm{R}_1,\bm{R}_1}(\bm{q})   = U\,,\\[8pt]
V^{P,0}_{\bm{R}_l,\bm{R}_l}(\bm{q})&= V^{C,0}_{\bm{R}_l,\bm{R}_l}(\bm{q})= V_1\,,
\end{align}
for $l\in\{2,3,4,5,6,7\}$.
The expressions for the more remote interaction terms $V_2$ and $V_3$ read
\begin{align}
V^{P,0}_{\bm{R}_{l'},\bm{R}_{l'}}(\bm{q})&= V^{C,0}_{\bm{R}_{l'},\bm{R}_{l'}}(\bm{q})= V_2,
\end{align}
for $l'\in\{10,11,14,15,18,19\}$ and 
\begin{align}
V^{P,0}_{\bm{R}_{l''},\bm{R}_{l''}}(\bm{q})&= V^{C,0}_{\bm{R}_{l''},\bm{R}_{l''}}(\bm{q})= V_3\,,
\end{align}
for $l''\in\{8,9,12,13,16,17\}$. 
Finally, we obtain
\begin{align}
V^{D,0}_{\bm{R}_{1},\bm{R}_{1}}(\bm{q})=&\ U + V_1 \sum_{l} e^{i\bm{R}_l\bm{q}}\nonumber\\
&+ V_2 \sum_{l'} e^{i\bm{R}_{l'}\bm{q}} + V_3 \sum_{l''} e^{i\bm{R}_{l''}\bm{q}}\,. 
\end{align}
This completes our TUFRG scheme. For more details on the numerical implementation see App.~\ref{sec:loopintegration}.

\section{Technical details}\label{sec:loopintegration}

In the numerical evaluation of the TUFRG flow equations, the bubble integrations in Eqs.~\eqref{eq::bubble1} and~\eqref{eq::bubble2} are the bottleneck.
This is because of two reasons: 
(1)~The integration kernel of the bubbles will form sharp features in vicinity of the Fermi surface when the flow parameter reaches small values such that it becomes gradually more difficult to integrate the function for lower scales. 
(2)~The bubbles have to be evaluated for $N_{\bm{q}} \times N_l^2$ different combinations for momenta $\bm{q}$ and form factors $R_l$. 
Both aspects combined lead to a challenge in both quantity and quality as one has to define an integration scheme which resolves the emerging peaks correctly along the flow while staying computationally performant such that the $N_{\bm{q}} \times N_l^2$ different integrals per RG step do not extend the execution time to an unreasonable amount.

\subsection{Adaptive integration routine}

\begin{figure}[t!]
\includegraphics[width=\columnwidth]{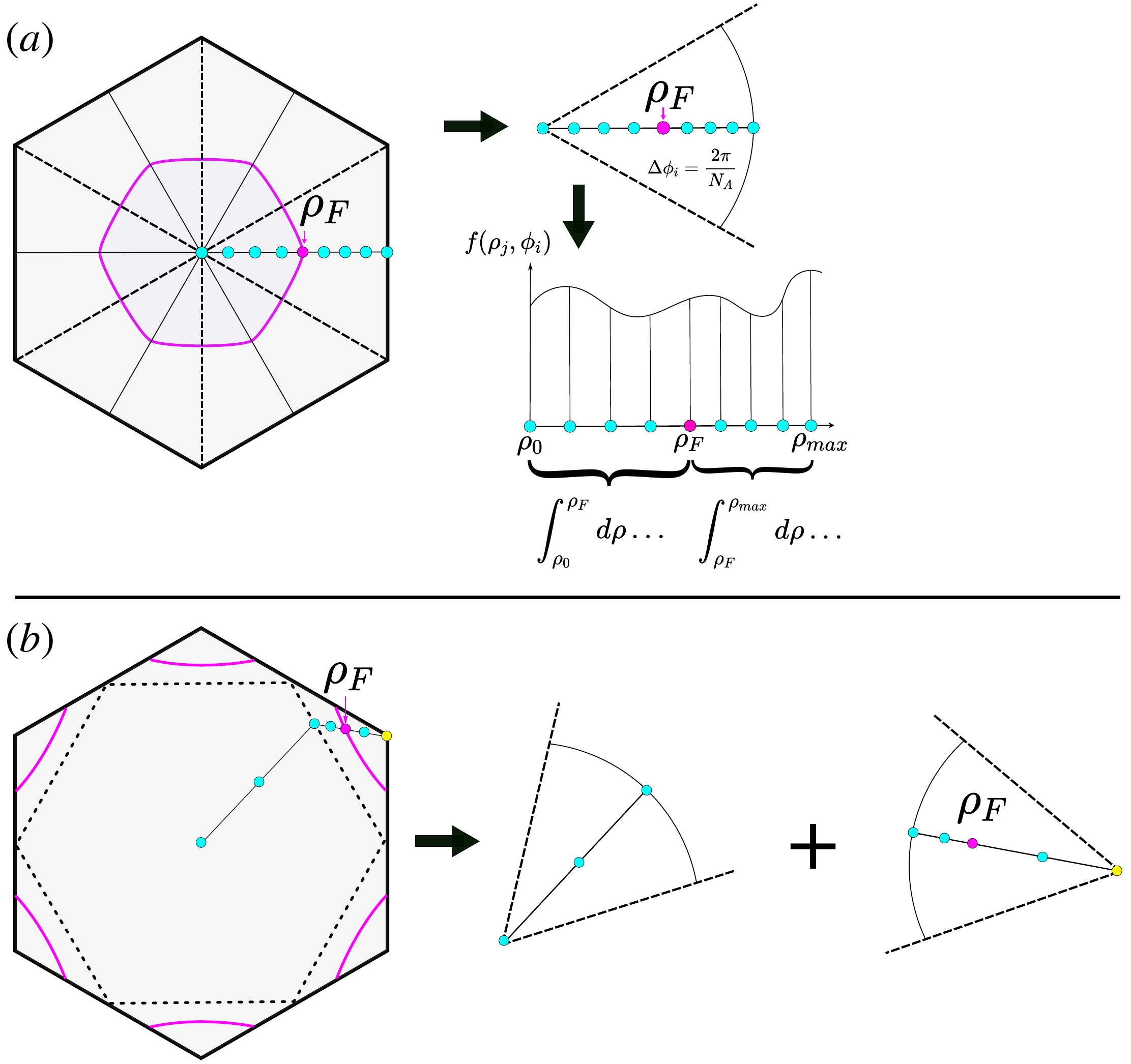}
\caption{\textbf{Schematics of the bubble integration routine.}
\textbf{(a)} 
Example with $N_R=5$ and $N_A=6$ for a single closed Fermi surface around the $\Gamma$ point.
The function is integrated along the six solid lines and the angular width for all lines is equal. 
Dashed lines mark the boundaries of the patches. 
The two integrations with $N_R=5$ discretization points are evaluated on the colored dots. 
\textbf{(b)} Modified intergration scheme for Fermi surface pockets around the $K,K'$ points. 
Example with $N_R=3$ of a single 1D integration. The general strategy for the integration is as described before, but now the 1D integration lines feature a kink on the nesting line (dashed hexagon). 
The first integral ranges from the origin to the nesting line. The other two integral intervals are integrated also in polar coordinates, but starting from the $K$-points. Again, the position of the Fermi surface is always chosen as a disretization point, leading to two integrals for this contribution: the first one from the $K$-point to the Fermi line, and the second one from the Fermi line to the nesting line.
}
\label{fig::integration}
\end{figure}

To tackle the problem~(1), we first discuss the case of closed Fermi lines around the $\Gamma$ point and then extend the routine to the case of Fermi pockets around $K, K'$. 
We introduce polar coordinates with the addition that the radius for each angle varies to correctly cover the hexagonal form of the Brillouin zone.
For a general function $f(\bm{k})$ this translates to
\begin{equation}
 \label{eq::RadialInt}
 \int dk f(\bm{k}) = \int_{0}^{2\pi}d\phi \int_{0}^{\rho_{\mathrm{max}}({\phi})} f(\bm{k}(\rho,\phi)) \cdot \rho .
\end{equation}
We then choose angular resolution $N_A$ and radial resolution $N_R$, such that $N_A$ straight lines are placed into the BZ from the origin, see Fig.~\ref{fig::integration}(a). 
The emerging slices have the same angular distance $\Delta \phi$ to each other. 
For each of these slices, we perform two one-dimensional adaptive trapezoidal-rule integrations over the radial direction $\rho$; 
one in the range $[0,\rho_F]$ and a second one in the range $[\rho_F,\rho_{\mathrm{max}}(\phi)]$, where $\rho_F$ is the radius lying directly on the Fermi surface. 
This choice was made such that $\rho_F$ is always a discretization point and therefore emerging sharp peaks at lower scales are always evaluated in the numerical integration. 
Since the integration of the triangular shaped area is approximated by a circular arc, we get a systematic error which becomes small for large $N_A$.
Then, Eq.~\eqref{eq::RadialInt} translates into
\begin{align}
 \int dk f(\bm{k}) &\approx \sum_{i=0}^{N_A -1 } \Delta \phi \left[ \int_{0}^{\rho_{\mathrm{max}}(\phi_i)}  f(\bm{k}(\rho,\phi_i)) \cdot \rho \ d\rho \right]  \nonumber \\
&=\sum_{i=0}^{N_A -1 } \Delta \phi \Biggl[ \left( \int_{0}^{\rho_{F}(\phi_i)}  f(\bm{k}(\rho,\phi_i)) \cdot \rho \ d\rho \right)   \nonumber \\ 
&\quad+ \left( \int_{\rho_{F}(\phi_i)}^{\rho_{\mathrm{max}}(\phi_i)}  f(\bm{k}(\rho,\phi_i)) \cdot \rho \ d\rho \right)     \Biggr]\,,
\end{align}
with $\phi_i = 2\pi i/N_A$ and $\Delta \phi = 2\pi/N_A$.

The two integrals over $\rho$ are then integrated adaptively as follows: Both integrals are equipped with $N_R$ discretization points leading to $N_R-1$ sub-intervals for both intervals. 
Therefore, the total amount of discretization points is always $2N_R$ (double counting the point on the Fermi-surface).
The spacing of the points for the two integration regions does not have to be equal, since the point of the Fermi surface $\rho_F$ does not necessarily divide the complete interval $[0,\rho_{\mathrm{max}}]$ in two equal intervals.  
The $N_R-1$ sub-intervals of the two intervals are the actual object where the adaptive routine is applied to. 
For the error we choose a relative tolerance of $10^{-3}$ and an absolute tolerance of $10^{-10}$.
Each of these sub-intervals is then integrated adaptively until this precision is met.

Since the adaptive routine of one sub-interval becomes unnecessarily expensive if only a fractional area of this sub-interval consists of sharp features, it is reasonable to also adjust the radial resolution for lower temperature scales. 
Starting with $N_R=3$, the amount of discretization points will double after a lower order of magnitude is reached, i.e.
\begin{align}
    N_R = 3\cdot 2^{\ceil{\log_{10}(|t|/T)}}\,.
\end{align}
The angular discretization $N_A$ is fixed through the flow since the sharp features should mainly develop along the $\rho$~axis which is therefore treated with more care. For the angular resolution we choose $N_A=96$. 
It is advantageous to select a number which is a multiple of 12 for $N_A$, such  that the angular dissection of the hexagon is concurring with the rotation and mirror symmetries of the hexagon.
For Fermi lines with pockets around the $K,K'$ points, we slightly modify this integration routine as described in Fig.~\ref{fig::integration}(b).

\subsection{Symmetry considerations}

To take care of the second point, which makes the numerical evaluation costly, we can exploit the spatial form of the form factors and their translational invariance.
Since the bubbles in Eqs.~\eqref{eq::bubble1} and~\eqref{eq::bubble2} only depend on two complex-valued plane waves, i.e. $f_l(\bm{q})$ and $f_l'^{*}(\bm{q})$, and one real function composed of $n_f'$ and $\xi(\bm{k})$, we find
\begin{align}
    \dot{B}(\bm{q})_{l,l'}^{(\pm)} = \left[ \dot{B}(\bm{q})_{l',l}^{(\pm)}      \right]^\ast\,.
    \label{eq::bubblecomplex}
\end{align}
Using this, we can reduce the numerical effort for a chosen~$\bm{q}$ by roughly a factor $1/2$ (not exactly as Eq.~\eqref{eq::bubblecomplex} does not provide an advantage for cases where $l=l'$).

To take advantage of translational invariance we now examine the product of form factors in more detail. 
The integration kernel of the bubbles, Eqs.~\eqref{eq::bubble1} and~\eqref{eq::bubble2}, depends on the product 
of form factors, i.e.
\begin{align}
f_l(\bm{p}) \times f_{l'}^{*}(\bm{p}) = e^{ i(\bm{R}_l - \bm{R}_{l'})\bm{p}} = e^{ i\bm{R}_{l-l'}\bm{p}}\,.
\label{eq::llform}
\end{align}
Therefore the bubbles depend on the real-space vector $\bm{R}_{l-l'}= \bm{R}_{l}-\bm{R}_{l'}$. 
We exploit this by identifying the combinations $(l,l')$ which result in the same vector to avoid re-calculating integral combinations $(l,l')$ with the same value.
Consider for example Fig.(\ref{fig::lattices}) where one can verify that
\begin{align}
(\bm{R}_2 - \bm{R}_5) = (\bm{R}_7 - \bm{R}_{13}) = (\bm{R}_8 - \bm{R}_7) = \bm{R}_{10}\,,
\end{align}
such that the bubbles $\dot{B}(\bm{q})_{l,l'}^{(\pm)}$ result in the same value for all of these $(l,l')$ pairings. 
We can therefore also relabel the bubble as:
$\dot{B}(\bm{q})_{l,l'}^{(\pm)}=\dot{B}(\bm{q})_{l-l'}^{(\pm)}$, depending on only one plane wave form factor with real-space vector $\bm{R}_{l-l'}$, see Eq.~\eqref{eq::llform}.
To use these relations systematically, we distinguish two different form factor shells: (1)~shells that are defined by including all form factors of the same spatial distance from the origin, i.e. the neighbor shells and (2)~shells with equal hexagonal distance, i.e. shells which include all form factors which are placed on the $N_s$-th hexagon of the triangular lattice, cf. Fig.~\ref{fig::lattices}. 
For this discussion, we call the $i$-th neighbor shell $S_n(i)$ and the $i$-th hexagonal-distance shell $S_{\mathrm{HD}}(i)$.

Considering the geometry we see that taking the $M$-th neighbor shell into account, all combinations of $(\bm{R}_l - \bm{R}_{l'})$ lie in a hexagonal-distance shell between $S_{\mathrm{HD}}(0)$ and $S_{\mathrm{HD}}(2M)$. 
To obtain a numerical advantage for calculating all entries of  $\dot{B}(\bm{q})_{l,l'}^{(\pm)}$ for a chosen $q$ we then apply the following procedure:
\begin{itemize}
  \item[--] Choose the amount of included form factors such that the largest reached hexagon is completely filled. 
  The highest filled hexagonal-distance shell is called $M$: $S_{\mathrm{HD}}(M)$, e.g., up to the third-neighbor shell which corresponds the form factors up to the second hexagonal-distance shell, $M=2$.
  \item[--] Calculate $\dot{B}(\bm{q})_{l-l'}^{(\pm)}$ for all $(\bm{R}_l - \bm{R}_{l'})$ included in $S_{\mathrm{HD}}(2M)$. 
  Using relation Eq.~\eqref{eq::bubblecomplex} we perform half of these calculations and obtain the other half by complex conjugation. 
  The number of calculations is then
  \begin{align}
  \hspace{1cm} \#\mathrm{calculations} &= \sum_{i=0}^{2M} S_{\mathrm{HD}}(i) = 1+\frac{1}{2}\sum_{i=1}^{2M} S_{\mathrm{HD}}(i) \nonumber \\
   &=1+3M+6M^2 \nonumber
  \end{align}
  \item[--] Finally the results of $\dot{B}(\bm{q})_{l-l'}^{(\pm)}$ can easily be related to the corresponding entries $\dot{B}(\bm{q})_{l,l'}^{(\pm)}$.
\end{itemize}
In the example $M=2$ one gets 19 form factors, cf. Fig.~\ref{fig::lattices}, resulting in $19 \times 19=361$ entries for $\dot{B}(\bm{q})_{l,l'}^{(\pm)}$ for a chosen $\bm{q}$. 
Combining the symmetry Eq.~\eqref{eq::bubblecomplex} with the approach described above, this reduces the number of integration to $1+3\cdot(2)+6\cdot(2)^2= 31$, providing an overall numerical speedup of factor $\sim 10$. 

\subsection{Differential equation solver}

The flow equations Eqs.~\eqref{eq::ffflowP}-\eqref{eq::ffflowD} are solved numerically by using an adaptive forward Euler method. 
For example the $P$-channel is evolved 
as
\begin{align}
    P^{l,l'}_{i+1}(\bm{q}) = P^{l,l'}_{i}(\bm{q}) + \frac{d}{d\Lambda}P^{l,l'}(\bm{q}) \cdot d\Lambda_{i}
\end{align}
where for $\frac{d}{d\Lambda}P^{l,l'}(\bm{q})$ the r.h.s of Eq.~\eqref{eq::ffflowP} is used. 
The stepsize $d\Lambda_{i}$ is adapted if the RG scale drops below a threshold or if the increment $\frac{d}{d\Lambda}P^{l,l'}(\bm{q})$ increases beyond a threshold.
The latter criterion may indicate the onset of a divergence where the differential equation behaves stiffly. 
Explicitly, our combined criteria are
\begin{align}
d\Lambda_{i+1}=&\min \Big( \frac{1}{20}\Lambda_i, \frac{1}{20}\max |\frac{d}{d\Lambda}P^{l,l'}(\bm{q})|, \nonumber \\ 
&\frac{1}{20}\max |\frac{d}{d\Lambda}C^{l,l'}(\bm{q})| ,\frac{1}{20}\max |\frac{d}{d\Lambda}D^{l,l'}(\bm{q})|            \Big) \nonumber \\
    \Lambda_{i+1} =& \Lambda_{i} - d\Lambda_{i+1} .\nonumber 
\end{align}
The initial scale $\Lambda_0$ is chosen to be equal to the bandwidth $W$ of the model. 
The solver stops the flow either when the absolute value of one entry of the $P,C$ or $D$ channel has surpassed the triple value of the bandwidth or when $\Lambda_i$ becomes smaller than $\Lambda_{\mathrm{stop}}/t=10^{-6}$.

\subsection{Vertex symmetries}

\begin{figure}[t!]
\centering
    \includegraphics[width=0.95\columnwidth]{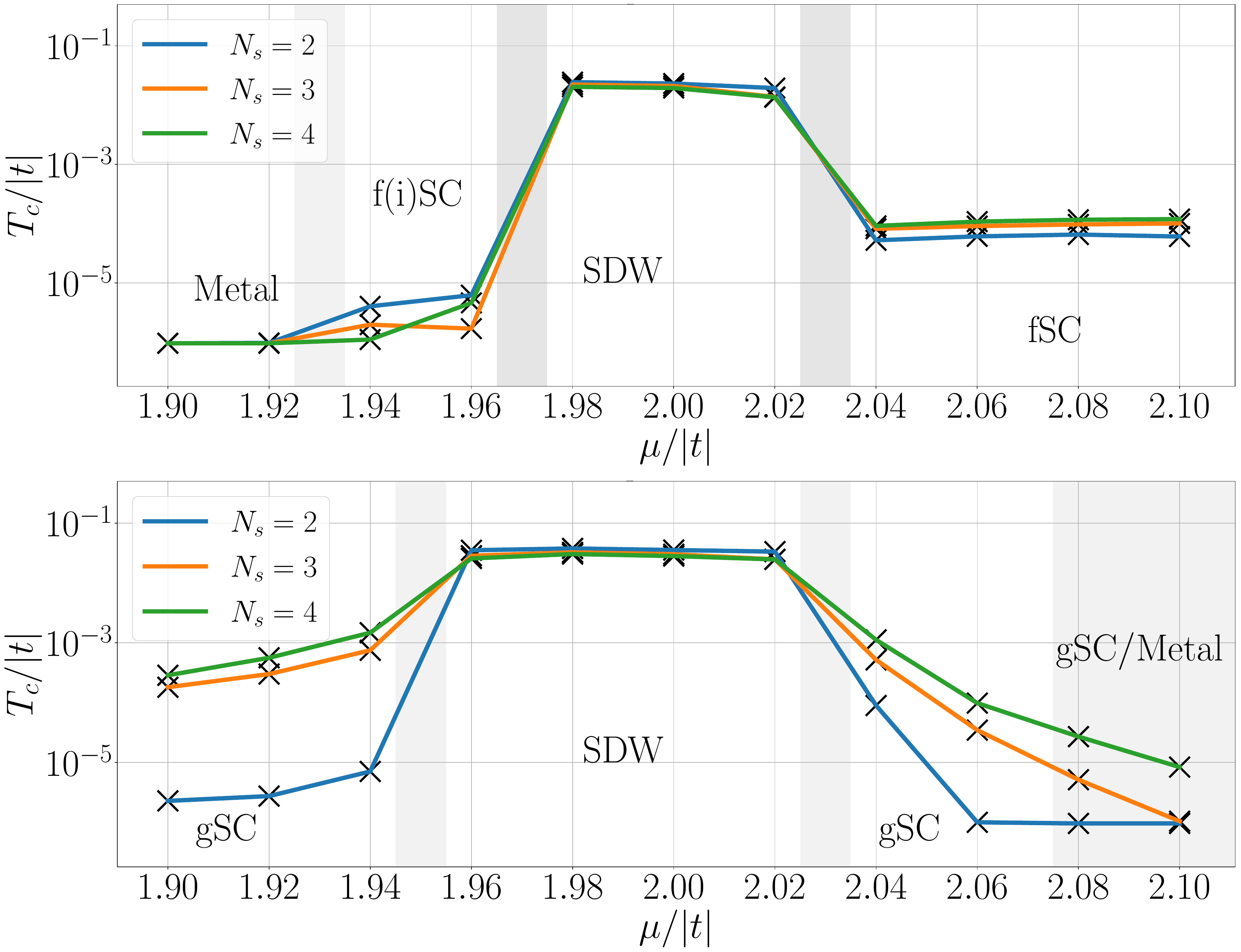}
    \caption{\textbf{Convergence checks for} $U=4t_1$ and $V_1=0$ (top) or $V_1=t$ (bottom) with $N_q=540$. 
    Instability types are labeled and phase boundaries are indicated by gray shadings.}
    \label{fig::convergence8}   
\end{figure}

As previously discussed in Ref.~\cite{PhysRevB.99.245140}, the symmetries of the lattice can be exploited to reduce the numerical effort for calculating the flow equations by a factor of twelve. 
The general symmetry relation for $X \in \{P,C,D\}$ is
\begin{align}
    X^{l,l'}(\bm{q}) = X^{Q\bm{R}_l,Q\bm{R}_{l'}}(Q\bm{q})\,,
\end{align}
where $Q$ is a symmetry operation allowed by the lattice. 
Note that the symmetry relations derived in Ref.~\cite{PhysRevB.99.245140} also include orbital degrees of freedom which are not present in our application. 
We choose two symmetry relations:
(1)~$Q=M_x$ , i.e. the mirror plane along the $x$~axis through the $\Gamma$ point and (2)~$Q=R_{2\pi/6}$, i.e. a rotation by angle $2\pi/6$.
Using these two operations it is sufficient to carry out the calculations for only $1/12$ of the momenta $\bm{q}$ in the flow equations.
We have also indicated this in the top panel of Fig.~\ref{fig::convergenceN20}. 
Exploiting the mirror and rotational symmetries, we eventually obtain all values for $P^{l,l'}(\bm{q})$, $C^{l,l'}(\bm{q})$ and $D^{l,l'}(\bm{q})$. 
In most of our calculations we use $N_{\bm{q}} = 540$ and consequently we only evaluate $540/12= 45$ entries while the rest is obtained by symmetry.

\section{Convergence checks for $N_q=540$}
\label{app::shells}

We perform the equivalent convergence check for $N_q=540$ with respect to the number of form factor shells that is shown in the main text for $N_q=1092$. We find the same conclusions, i.e. the number of shells can be important at phase boundaries and for the extraction of pairing symmetries, see Fig.~\ref{fig::convergence8}. However, for most parameter choices the results converge both qualitatively and quantitatively.

\bibliography{TUFRG_Triangular}

\end{document}